%% file: Mbh-Lbt_II.tex
\documentclass[12pt,usenatbib,usegraphicx]{emulateapj}

\usepackage{epsfig}
\usepackage{latexsym,amssymb}
\usepackage{amsmath}
\usepackage{textcomp}
\usepackage{dcolumn}
\usepackage[stable]{footmisc}
\usepackage{booktabs}

\include{Mbh-Lbt_abbrev}

\lefthead{L\"asker et al.}
\righthead{\smbh\ scaling relations with NIR luminosities}
\slugcomment{Version \today; accepted by {\it The Astrophysical Journal}.}

\begin{document}

\title[The correlation with near-infrared luminosity revisited]
 {Supermassive Black Holes and Their Host Galaxies -- \\ \textrm{II}. The correlation with near-infrared luminosity revisited}

\author{Ronald L\"{a}sker$^{1,2}$, Laura Ferrarese$^2$, Glenn van de Ven$^1$, Francesco Shankar$^{3,4}$}
\affil{${}^1$ Max-Planck Institut f\"ur Astronomie, K\"onigstuhl 17, D-69117, Heidelberg, Germany; e-mail:laesker@mpia.de}
\affil{${}^2$ NRC Herzberg Institute of Astrophysics, 5071 West Saanich Road, Victoria, BC V9E2E7, Canada}
\affil{$^3$ GEPI Observatoire de Paris, CNRS, Univ. Paris Diderot, 5 place Jules Janssen, 92195 Meudon, France}
\affil{$^4$ Department of Physics and Astronomy, University of Southampton, Highfield, SO17 1BJ}

\begin{abstract}
We present an investigation of the scaling relations between Supermassive Black Hole (\smbh) masses, $\mbh$, and their host galaxies' $K$-band bulge ($\lbul$) and total ($\ltot$) luminosities. The wide-field WIRCam imager at the Canada-France-Hawaii-Telescope (CFHT) was used to obtain the deepest and highest resolution near infrared images available for a sample of 35 galaxies with securely measured $\mbh$, selected irrespective of Hubble type. For each galaxy, we derive bulge and total magnitudes using a two-dimensional image decomposition code that allows us to account, if necessary, for large- and small-scale disks, cores, bars, nuclei, rings, envelopes and spiral arms. We find that the present-day $\mbh-\lbul$ and $\mbh-\ltot$ relations have consistent intrinsic scatter, suggesting that $\mbh$ correlates equally well with bulge \textit{and} total luminosity of the host. Our analysis provides only mild evidence of a decreased scatter if the fit is restricted to elliptical galaxies. The log-slopes of  the $\mbh-\lbul$ and $\mbh-\ltot$ relations are  $0.75\pm0.10$ and $0.92\pm0.14$, respectively. However, while the slope of the $\mbh-\lbul$ relation depends on the detail of the image decomposition, the characterization of $\mbh-\ltot$ does not. Given the difficulties and ambiguities of decomposing galaxy images into separate components, our results indicate that $\ltot$ is  more suitable as a tracer of \smbh\ mass that $\lbul$, and that the $\mbh-\ltot$ relation should be used when studying the co-evolution of \smbh s and galaxies.
\end{abstract}

\section{Introduction}
\label{sec:intro}

The correlation between Supermassive Black Hole (\smbh) masses, $\mbh$, and the luminosities of their host galaxies' bulges, $\lbul$, is significant for at least two reasons.

First, there is to this date no obvious nor unique theoretical framework that satisfactorily explains its existence and characteristics  (see, e.g., \citealt{Shankar+12} and the discussion in \citealt{Kormendy+Ho13}). Understanding this, and other correlations between \smbh s and host galaxy properties, has important implications for understanding the origin of \smbh s and, possibly, the evolution of galaxies (e.g. \citealt{SilkRees98,Cattaneo+06,Lusso+Ciotti11}). Models of \smbh-galaxy co-evolution include mergers of galaxies and their \smbh s \citep[e.g.][]{Peng07}, but typically also invoke an interaction (``feedback'') between \smbh s and their hosts \citep[e.g.][]{Granato_etal04, Hopkins_etal06, Croton_etal06, Shankar_etal06, VolNatGul11b, Fanidakis+12}. The relative importance of mergers and feedback, their timescales, efficiencies and frequencies, may be constrained by comparing model predictions with the observed $\mbh-\lbul$ relation.

Second, the $\mbh-\lbul$ relation allows to infer \smbh\ masses, which are difficult to measure directly, from observationally much more accessible luminosities. This is used, for example, in the context of establishing the black hole mass function (BHMF) and its redshift evolution. Furthermore, the $\mbh-\lbul$ relation derived from dynamically measured \smbh\ masses is also employed to calibrate methods of measuring $\mbh$ in active galaxies, namely reverberation mapping and secondary methods that are based on it \citep[e.g.][]{Onken_etal04,FerrareseFord05,McGill_etal08}. The latter are frequently used to investigate the redshift evolution of \smbh\ masses and \smbh-galaxy scaling relations \citep[e.g.,][]{Bennert_etal11}.

The $\mbh-\lbul$ relation shares both theoretical significance and predictive power with other correlations between \smbh\ mass and host galaxy properties, most notably  the bulge stellar velocity dispersion, $\sigma_\star$, as well as stellar and dynamical mass, $M_\mathrm{\star(,bul)}$ and $M_\mathrm{dyn(,bul)}$. Of special interest in this context is the correlations' intrinsic scatter, $\epsilon$, since a relation with low scatter allows a more precise estimate of $\mbh$, provided that the correlating quantity is observationally accessible with adequate precision. Yet, aside from desiring a ``tight'' relation as predictive tool, it is worthwhile to note that from a theoretical perspective, all correlations, not only the one with the lowest scatter, are of interest, since they serve as an additional constraint to evolutionary models. Consequently, it is imperative for the calibration of these relations to be as precise and unbiased as possible. For example, the models of \cite{JahnkeMaccio11}, where merging is the main driver behind the emergence of the $\mbh-M_\star$ relation at $z=0$, predict a lower scatter than observed unless star formation, black-hole accretion and disk-to-bulge conversion are included.

Unfortunately, the present characterization of the $\mbh-\lbul$ relation and its intrinsic scatter are not secure. Published parameter values range from $\sim0.9$ to $\sim1.3$ for the log-slope, and $\sim0.3$ to $\sim0.6$ for the scatter $\eps$, with a typical $1\sigma$-uncertainty of $\sim10\%$. Of particular interest is the relation at near-infrared (NIR) wavelengths, since, compared to optical, NIR luminosities are a better tracer of stellar mass and are less affected by dust extinction. The first study of $\mbh-L_\mathrm{bul,NIR}$ relations was based on \twomass\ $J$, $H$ and $K$-band data \citep[][MH03 hereafter]{MH03}. The authors estimated the scatter in the relation, $\eps$, to be between 0.5 and 0.3, the former obtained using the full sample of $\mbh$ available at the time, the latter using a subsample of 27 galaxies deemed to have {\it reliable} estimates of $\mbh$. Surprisingly, MH03 found the intrinsic scatter in the NIR to be equivalent to the scatter in the optical ($B$-band) for the same sample, despite the expected advantages (less sensitivity to stellar population and dust effects) of working at longer wavelengths. Given the low resolution and depth of the \twomass\ images, uncertainties in the bulge magnitudes derived from image decompositions can easily be underestimated, which would lead to overestimate the intrinsic scatter in the relation. This issue therefore demands further attention.

More recently, \citeauthor{V12} (\citeyear{V12}, hereafter V12) presented an updated relation using data from the \ukidss\ survey. The improved spatial resolution and depth of the data allowed them to account for nuclei, bars and cores when modeling galaxy images (while MH03 restricted their analysis to a bulge+disk decomposition). Based on a sample of 25 (19) galaxies, V12 found a characterization of the relation in agreement with MH03 \citep[see][]{Graham07}, but noted that the scatter increased to $~0.5\dex$ ($~0.4\dex$ for their restricted sample excluding barred galaxies), in spite of the clearly improved quality of the photometric data and analysis.

To further investigate the $K$-band $\mbh-\lbul$ relation, we used the WIRCam instrument at CFHT to conduct a dedicated imaging survey, resulting in the deepest currently available homogeneous NIR-data set of known \smbh\ host galaxies with reliable direct $\mbh$ measurements. Since our goal is to perform an accurate decomposition of each galaxy into separate morphological components, we required $\lesssim 1\arcsec$ image resolution. In order to further reduce photometric uncertainties, we placed special emphasis on developing an observation strategy and data reduction pipeline that reliably accounts for the high and variable NIR background. Using the latest \galfit\ profile fitting code \citep[\gfthree,][]{GF3}, we find that most galaxies require additional components in addition to a canonical bulge and (in most cases) disk, and demonstrate the large impact of the decomposition details on the derived bulge magnitudes. We additionally characterize the relation between $\mbh$ and total galaxy luminosity, $\ltot$,  and compare it with the $\mbh-\lbul$ relation.

Details of the data acquisition and analysis, as well as a detailed discussion of each galaxy, are presented in a companion paper (L\"{a}sker et al. 2013, hereafter Paper-I). The present paper focuses on the derivation of the $\mbh-L$ scaling relations and is organized as follows: In \S\ref{sec:datalum} we describe the sample selection and summarize the analysis leading to our bulge and total luminosity measurements. \S\ref{sec:results} presents the ensuing scaling relations, including comparison with previous publications. We discuss our results in \S\ref{sec:discussion} and draw conclusions in \S\ref{sec:summary}. We elaborate on the correlation fitting method and the treatment of data uncertainties in the appendix.

\section{Bulge and total luminosities from superior near-infrared imaging}
\label{sec:datalum}

\subsection{Sample of \smbh\ host galaxies}
\label{sec:sample}

Our sample comprises all galaxies visible from Mauna Kea for which a secure estimate of $\mbh$ had been published at the time our observing proposal was submitted (March 2008), regardless of  Hubble type. ``Reliable" $\mbh$ estimates are those derived based on dynamical modeling of the spatially resolved kinematics of gas or stars. We do not consider targets where only upper limits on $\mbh$ are available. {In order to avoid systematic uncertainties implied by possibly unjustified model assumptions,} we exclude galaxies for which $\mbh$ was derived from gas kinematics based on models with fixed disk inclination (e.g. NGC4459 and NGC4596, \citealt{Sarzi+01}), or from stellar kinematics based on isotropic dynamical models (NGC4594, \citealt{Kormendy88}; NGC4486b, \citealt{Kormendy+97}). Also excluded are measurements flagged as being particularly uncertain in the original papers (M31, \citealt{Bacon+01}; NGC1068 \citealt{Greenhill+96}; Circinus \citealt{Greenhill+03}, and the unpublished \smbh\ mass of NGC4742 \citep[initially cited in][]{T02}. NGC5128 (CenA) has been excluded because of declination constraints. Where available, we have adopted updated BH mass measurements that account for the DM halo \citep[e.g.][]{SchulzeGebhardt11}, triaxiality \citep[e.g.][]{vdBosch_deZeeuw10} and velocity dispersion of the gas disk \citep[e.g.][]{Walsh+10} in the model.

With respect to the work of MH03, we did not observe seven galaxies in their sample (M31, NGC1068, NGC4459, NGC4594, NGC4596, NGC4742 and NGC5128), although we note that $\mbh$ for four of these galaxies was also deemed of uncertain quality  by MH03, and not included in their final analysis. Compared to V12, we did not observe five galaxies that are part of their final sample (NGC1068, NGC2960, NGC4303, NGC4459 and NGC4596), as well as four additional galaxies (NGC863, NGC4435, NGC4486b and UGC9799) that V12 also excluded from the fit as having uncertain $\mbh$. 
On the other hand, our study includes several galaxies that were not part of MH03's nor V12's analysis. Compared to MH03, we include seven galaxies for which $\mbh$ became available after their paper was published (IC4296, NGC1300, NGC1399, NGC2748, NGC3227, NGC3998, and PGC49940==A1836-BCG). None of these galaxies were part of the sample of V12; in addition, our sample contains 14 galaxies that were not analyzed by V12 (CygA, IC1459, NGC821, NGC1023, NGC2787, NGC3115, NGC3377, NGC3379, NGC3384, NGC3608, NGC4291, NGC5252, NGC6251 and NGC7457). Finally, we of course did not obtain a new measurement of the  Milky Way bulge magnitude, and this galaxy is therefore not part of our final sample, in spite of its exceedingly well measured $\mbh$.

The above selection results in the sample of 35 galaxies listed in Table \ref{tab:targets}. The distances listed in the table are measured using Surface Brightness Fluctuations when available\footnote{A constant of $0.06\mg$ has been subtracted from the distance moduli as given in \cite{Tonry_etal01} to account for the updated Cepheid distances presented in \cite{Freedman_etal01}.} or otherwise estimated from the systemic velocity assuming $H_0=72\pm8\,\textrm{km/s}$, \citep[][]{Freedman_etal01}. All \smbh\ masses, which are taken from the literature, have been rescaled to the new adopted distances. Galactic foreground extinctions are taken from the NED database. $K$-band luminosities, as derived from our images, are also given (see Paper-I).

\begin{table*}
 \caption{\smbh\ host galaxy sample}
 \newcolumntype{d}[1]{D{#1}{#1}{-1}}
 \centering
 \begin{tabular}{lcd{.}d{.}d{/}d{.}d{.}d{-}}
  \toprule
  Galaxy & Hubble type & \multicolumn{1}{c}{D [Mpc]} & \multicolumn{1}{c}{$\mbh~[10^8\,\msun]$} & \multicolumn{1}{c}{$\Delta\mbh~(+~/~-)$} & \multicolumn{1}{c}{$L_K~[10^{11}\,\msun]$} & \multicolumn{1}{c}{$R_e~[\mathrm{kpc}]$} & \multicolumn{1}{c}{Ref.} \\
  (1) & (2) & \multicolumn{1}{c}{(3)} & \multicolumn{1}{c}{(4)} & \multicolumn{1}{c}{(5)} & \multicolumn{1}{c}{(6)} & \multicolumn{1}{c}{(7)} &  \multicolumn{1}{c}{(8)} \\
 \midrule
  \input{table_targets}
 \bottomrule
\end{tabular}
 \tablecomments{Columns (1) and (2) give the name and Hubble type of each galaxy. Distances given in column (3) are based on surface brightness fluctuations (SBF) whenever available. A constant $0.06\mg$ has been subtracted from the distance moduli taken from Tonry et al. to account for the updated Cepheid distances presented in \protect\cite{Freedman_etal01}. Redshift distances are derived from the velocities reported in NED, corrected for Virgocentric infall following \protect\cite{HubbleKey_XXVIII}, and adopting a value of the Hubble constant of $72 \pm 8$ km/s/Mpc \protect\citep[][]{Freedman_etal01}, which is consistent with the Cepheid-based calibration of the SBF distances used for much of the sample. Our adopted \smbh\ masses and errors are given in columns (4) and (5). Columns (6) and (7) list galaxy luminosities and effective radii as derived from our WIRCam data via aperture photometry. Finally, column (8) gives the references. Distance references (first digit) are 1: \protect\cite{Tonry_etal01}; 2: redshift distances (NED); 4: \protect\cite{Herrnstein_etal99}; 5: \protect\cite{Mei_etal07}; 7: \protect\cite{Jensen_etal03}; and 8: \protect\cite{Blakeslee_etal09}. \smbh\ masses (second digit) are 1: \protect\cite{Tadhunter_etal03}; 2: \protect\cite{Cappellari_etal02}; 3: \protect\cite{DallaBonta_etal09}; 4: \protect\cite{Verolme_etal02}; 5: \protect\cite{SchulzeGebhardt11}; 6: \protect\cite{Bower_etal01}; 7: \protect\cite{Atkinson_etal05}; 8: \protect\cite{Gebhardt_etal07}; 9: \protect\cite{Sarzi_etal01}; 10: \protect\cite{EmsellemDejongheBacon99}; 11: \protect\cite{Davies_etal06}; 12: \protect\cite{Barth_etal01}; 13: \protect\cite{vdBosch_deZeeuw10}; 14: \protect\cite{Walsh+12}; 15: \protect\cite{Herrnstein_etal05}; 16: \protect\cite{FerrareseFordJaffe96}; 17: \protect\cite{Cretton_vdBosch96}; 18: \protect\cite{Walsh+10}; 19: \protect\cite{Gebhardt_etal11}, but see \protect\cite{Walsh+13} for a different estimate; 20: \protect\cite{ShenGebhardt10}; 21: \protect\cite{Capetti_etal05}; 22: \protect\cite{FerrareseFord99}; 23: \protect\cite{vdMarel_vdBosch98}.}
\label{tab:targets}
\end{table*} 

\subsection{Data analysis and galaxy decomposition}
\label{subsec:datadecomp}

Full details of the data reduction and analysis are given in Paper-I. All data were obtained with WIRCam at the Canada-France-Hawaii-telescope, have subarcsecond resolution (the median seeing on the final image stacks is $0\farcs8$, a factor of 2-3 improvement over \twomass), and signal-to-noise ratio of S/N=1 at $24\textrm{mag/arcsec}^2$, 4 and 2 mag deeper than \twomass\ and \ukidss, respectively. The observing strategy and data reduction were optimized to reduce both random and systematic uncertainty when subtracting the bright and highly variable NIR background.

Our measurements of apparent magnitudes are based on two-dimensional (2D) image decomposition performed with \gfthree\ \citep[][]{GF3}, the details of which are given in Section 2 of Paper-I, which also includes a table listing bulge parameters and a discussion of the uncertainties associated with the models. To summarize,  we require each galaxy model to contain a bulge  with a \sersic\ radial surface brightness profile \citep[][]{Sersic63}. When warranted by the data, we also add a disk  with exponential radial profile (equivalent to a \sersic\ profile with index $n\equiv 1$). Such \sersic\ bulge (+ exponential disk) models have been applied in most previous studies, and we refer to them as ``standard'' models.

After fitting all images with standard models, and measuring the corresponding bulge and total magnitudes, most (30 out of 35) galaxies showed characteristic residuals in the model-subtracted images. While large residuals are expected for spiral galaxies, we observe them in \textit{all} galaxies with a disk component, and even in some ellipticals. This led us to attempt more detailed and complex models that include additional components (bars, nuclei, inner disks, envelopes, spiral arms), necessary profile modifications (such as diski-/boxiness and truncations) and masking the cores of giant ellipticals. We refer to these more complex models (i.e. including components in addition to bulge and disk) as ``improved'' models. To decide whether to include additional components as well as profile modifications in our analysis, we take a multi-prong approach, based on an examination of the radial profiles along the semimajor axis (obtained by \iraf\ {\it ellipse}), on the $\chi^2$ of the fit, and on a visual analysis of the residual image. We also verified the suitability of the fits by relaxing the assumption of an exponential profile for the disks. We required the models to be non-degenerate, and to converge to a best-fit with a stable minimum-$\chi^2$ that does not depend on the choice of initial parameters.

\subsection{Bulge and total luminosities}
\label{subsec:luminosities}

An overview of the different definitions for the luminosities derived by applying standard and improved model decompositions is given in Table \ref{tab:lumdef}.

\begin{table*}
 \centering
 \caption{Definitions of Luminosities}
 \begin{tabular}{lll}
  \toprule
  Luminosity & Short name & Definition \\
  \midrule
  $\lbstd$ & ``standard bulge'' & bulge component luminosity from a standard \sersic-bulge(+exponential disk) model \\
  $\lbmin$ & ``minimal bulge'' & bulge component luminosity from an improved (additional components or masked core) model  \\
  $\lbmax$ & ``maximal bulge'' & total minus disk and (if present) spiral arm luminosity of the improved model \\
  $\boldsymbol \lsph$ & \textbf{``spheroid''} & \textbf{luminosity of bulge component plus envelope (if present) in the improved model} \\
  \midrule
  $\lser$ & ``\sersic'' & luminosity of single-\sersic\ models (all galaxies) \\
  \midrule
  $\ltstd$ & ``standard total'' & sum of bulge and disk luminosity of a standard model \\
  $\boldsymbol \ltimp$ & \textbf{``improved total''} & \textbf{sum of all component luminosities from an improved model} \\
  $\liso$ & ``isophotal'' & luminosity within aperture delimited by the $24\magarcsec$ isophote \\
  \midrule
  $L_{\{\cdot\}}\ellip$ & ``ellipticals'' & luminosities of elliptical galaxies \\
 \bottomrule
\end{tabular}
 \tablecomments{Summary of the luminosities used in this paper. The method of obtaining standard and improved models is presented in Section \ref{sec:datalum} and  detailed in Paper-I. }
\label{tab:lumdef}
\end{table*} 

For all galaxies, we measure $\lbstd$, the luminosity of the bulge component in a ``standard" bulge(+disk) model. For galaxies requiring components in addition to a bulge and disk, we derive three distinct improved bulge luminosities: the ``minimal'' bulge luminosity ($\lbmin$) is the luminosity of the improved model's bulge component alone; the ``maximal'' bulge luminosity is derived by summing the flux of \textit{all} components \textit{except} the disk and, if present, spiral arms. Therefore, $\lbmin$ constitutes a lower, and $\lbmax$ an upper limit estimate for the bulge luminosity. $\lbmin$ and $\lbmax$ sometimes differ considerably from one another, as well as from $\lbstd$. For several galaxies, the latter is outside of the range defined by $\lbmin$ and $\lbmax$, indicating that a blind bulge+disk decomposition can significantly bias the results. A final estimate of the bulge luminosity, the ``spheroid luminosity'' ($\lsph$), includes the flux of the envelope component, in addition to the bulge component (see \S3.3 of Paper-I). As detailed in Paper-I, interpretation of the envelope is not straightforward, but it most likely represents part of the \textit{spheroidal} stellar distribution, which is not well fit by a single 2D-\sersic\ profile. This of course applies only to galaxies in which such an envelope is identified and required to obtain a suitable fit; for all other targets, $\lsph=\lbmin$.  

Combining the luminosities of all components provides the total luminosity, $\ltstd$ and $\ltimp$, of the standard and improved model, respectively. Naturally, for elliptical galaxies (fit by a single \sersic\ profile), the total equals the bulge luminosity. If the galaxy has a core (defined as a deficit of light relative to the inner extrapolation of the \sersic\ law that best fits the outer parts of the profile, e.g. \citealt{Ferrarese_etal06b}), the core is masked and the total luminosity obtained is $\ltimp$; for all other ellipticals, also $\ltimp=\ltstd$. In order to examine the effect of omitting any sub-components (including the exponential disk) from the model, as may occur in circumstances of insufficient image depth or resolution, we supply also the (total) luminosity from fitting single \sersic\ profile to \textit{all} galaxies ($\lser$). In addition, we derive a non-parametric luminosity from aperture photometry. The resulting isophotal luminosity, $\liso$, accounts for the flux inside $R_{24}$, the radius at which the surface brightness drops below $24\magarcsec$. Differences between $\ltstd$, $\ltimp$ and $\liso$ are relatively small compared to the variance between the various definitions of bulge luminosities.
                                                                                                                                                                                                                                                                                                                                                                                                                                                                                                                        
Although standard models rarely provide a good description for the data, they are useful to investigate the impact that the modeling complexity has on the BH scaling relations. Of course, in galaxies where a bulge(+disk) model suffices and no core is present (1 lenticular and 8 elliptical galaxies for our sample), we adopt luminosities from the standard model throughout. All luminosities are given in units of solar luminosity and have been converted from the absolute magnitudes given in Table 2 of Paper-I using the absolute $K$-band magnitude of the Sun\footnote{as taken from: \url{http://www.ucolick.org/~cnaw/sun.html}}, $M_{K,\odot}=3.28$.

In summary (see Table \ref{tab:lumdef}), we have four estimates for the bulge luminosity, $\lbul=\{\lbstd,~\lbmin,~\lbmax,~\lsph\}$, and four estimates for the total luminosity, $\ltot=\{\lser,~\ltstd,~\ltimp,~\liso\}$. $\lbstd$ and $\ltstd$ are derived using a ``standard" model (bulge plus, if needed, disk), while $\lbmin$, $\lbmax$, $\lsph$ and $\ltimp$ are derived from improved models that include additional components. By contrast, $\liso$ is a parameter-free measurement of total luminosity. 

\begin{table*}
 \centering
 \caption{Luminositiy values}
 \begin{tabular}{l*{13}c}
  \toprule
   Galaxy & class. & $\log\mbh$ & $\ltwom$ & $\liso$ & $\lser$ & $\ltstd$ & $\boldsymbol{\ltimp}$ & $\lbstd$ & $\lbmin$ & $\lbmax$ & $\boldsymbol{\lsph}$ & $\lmh$ \\
  \midrule
  \input{table_loglum}
  \bottomrule
 \end{tabular}
\tablecomments{Bulge and total luminosities used to produce the plots shown in Figures \ref{fig:mainfig}-\ref{fig:adopted}, and expressed as $\log (L/L_\odot)$. Also given are the logarithmic \smbh\ masses (with error bars symmetrized) and our galaxy classification (1: ''elliptical'', single \sersic\ profile / 2: ''lenticular'', disk present /  3: ''spiral'' galaxy, i.e. spiral component present), which is color-coded as red / green / blue in Figures \ref{fig:mainfig}, \ref{fig:mlbul} and \ref{fig:adopted}. Printed in boldface are the bulge and total luminosities used for our adopted scaling relations. All luminosities have been derived from our WIRCam data, except for $\ltwom$ (total luminosity as listed in the \twomass\ database) and $\lmh$ (derived from the bulge magnitudes listed in MH03), both of which were corrected to our distances. Aside from $\liso$, all luminosities are based on \galfit\ models (see \S\ref{subsec:datadecomp}). For a summary of the definitions used to derive luminosities in our study, see Table \ref{tab:lumdef} and Paper-I.}
\label{tab:loglum}
\end{table*} 

\section{Resulting scaling relations}
\label{sec:results}

For all sample galaxies, we plot the luminosities, $\logl_{\{\cdot\}}=\log\left(L_{\{\cdot\}}/\lsun\right)$ against \smbh\ mass, $\lmbh\equiv\log\left(\mbh/\msun\right)$, and fit a correlation with a ridgeline
\begin{equation}
 \lmbh = a + b(\logl_{\{\cdot\}}-11)~, \label{eqn:linrel}
\end{equation}
including a Gaussian\footnote{this choice of distribution is justified in \cite{G09}} intrinsic scatter $\epsbh$ in the direction of $\lmbh$ (y-axis) and $\epso$ in the direction perpendicular to the ridgeline. The subscript indicates one of the choices described in \S\ref{subsec:luminosities} and Table \ref{tab:loglum}. The corresponding $\lmbh$ and $\logl$ values are listed in Table \ref{tab:loglum}. We choose to pivot all relations around $L=10^{11}\lsun$ since this value is approximately equal to the mean luminosity of the sample, reducing the covariance between $a$ and $b$. In the remainder of this section, we  use the notation $\eps\equiv\epsbh$, as opposed to $\epso$. We also introduce the shorthand notation $\{p\}_{\{\cdot\}}\ellip$, where $p\in\{a,b,\eps\}$ may be any of the fit parameters, the subscript again indicates the type of luminosity measurement used, and the superscript (ell) is present if only elliptical galaxies were used in fitting the relation.

The parameter estimates are performed by calculating the three-dimensional posterior probability distribution of the parameters $(a,b,\eps)$ on a dense grid. The maximum of the likelihood function corresponds to the adopted parameters, while the quoted uncertainties correspond to the central 68\%-confidence interval of the marginalized distributions. We supplement our statistical analysis by estimates from bootstrap-resampling and find the results to be consistent. For more details on the choice of the fitting method, and the treatment of uncertainty in the data, please see Appendix \ref{app:fitmethod}. A summary of the parameters derived for all fitted scaling relations is given in Table \ref{tab:orig_full}.

\begin{table*}
 \centering
 \caption{Resulting scaling relations}
 \begin{tabular}{cl*{9}c}
  \toprule
   \multicolumn{2}{r}{Luminosity} & $a_0$ & $\Delta a$ & $b_0$ & $\Delta b$ & ${\epsbh}_{,0}$ & $\Delta\epsbh$ & ${\epso}_{,0}$ & $\sigma(\epso)$ & $r$ \\   
   \input{table_orig_full}
  \bottomrule
\end{tabular}
 \tablecomments{Best-fit parameters of the \smbh\ scaling relaions with luminosity, derived using our WIRCam data (Table 2 and eqn. \ref{eqn:linrel}). Column (1) indicates which luminosity was correlated with $\mbh$ (see Table \ref{tab:lumdef}). Columns (2), (4), and (6) contain the best-fit parameter values $(a,b,\epsbh)_0$ of each relation, as derived via the maximum-likelihood method. Columns (3), (5) and (7) give the relative location of the central 68\%(``1-$\sigma$'')-confidence interval, as derived from the posterior probability distribution. Column (8) gives the orthogonal intrinsic scatter, calculated from $b$ and $\epsbh$, and column (9) the standard deviation of its probability distribution as obtained via bootstrap resampling. Finally, column (10) indicates each relation's Pearson correlation coefficient. For more details about the fitting method, see appendix \ref{app:fitmethod}. Note that the confidence interval of $\epsbh$ is asymmetric with respect to the best-fit value because (i) after marginalizing the 3D-distribution over $a$ and $b$, $\epsbh$ closely follows a lognormal distribution, i.e. the median (and mean) are larger than the maximum, which is (ii) itself larger than ${\epsbh}_{,0}$ due to positive correlation of $\epsbh$ with $|a-a_0|$ and $|b-b_0|$.}
\label{tab:orig_full}
\end{table*} 

\subsection{Bulge versus total luminosity}
\label{subsec:mainres}

\begin{figure*}
  \begin{center}
    \includegraphics[width=18cm]{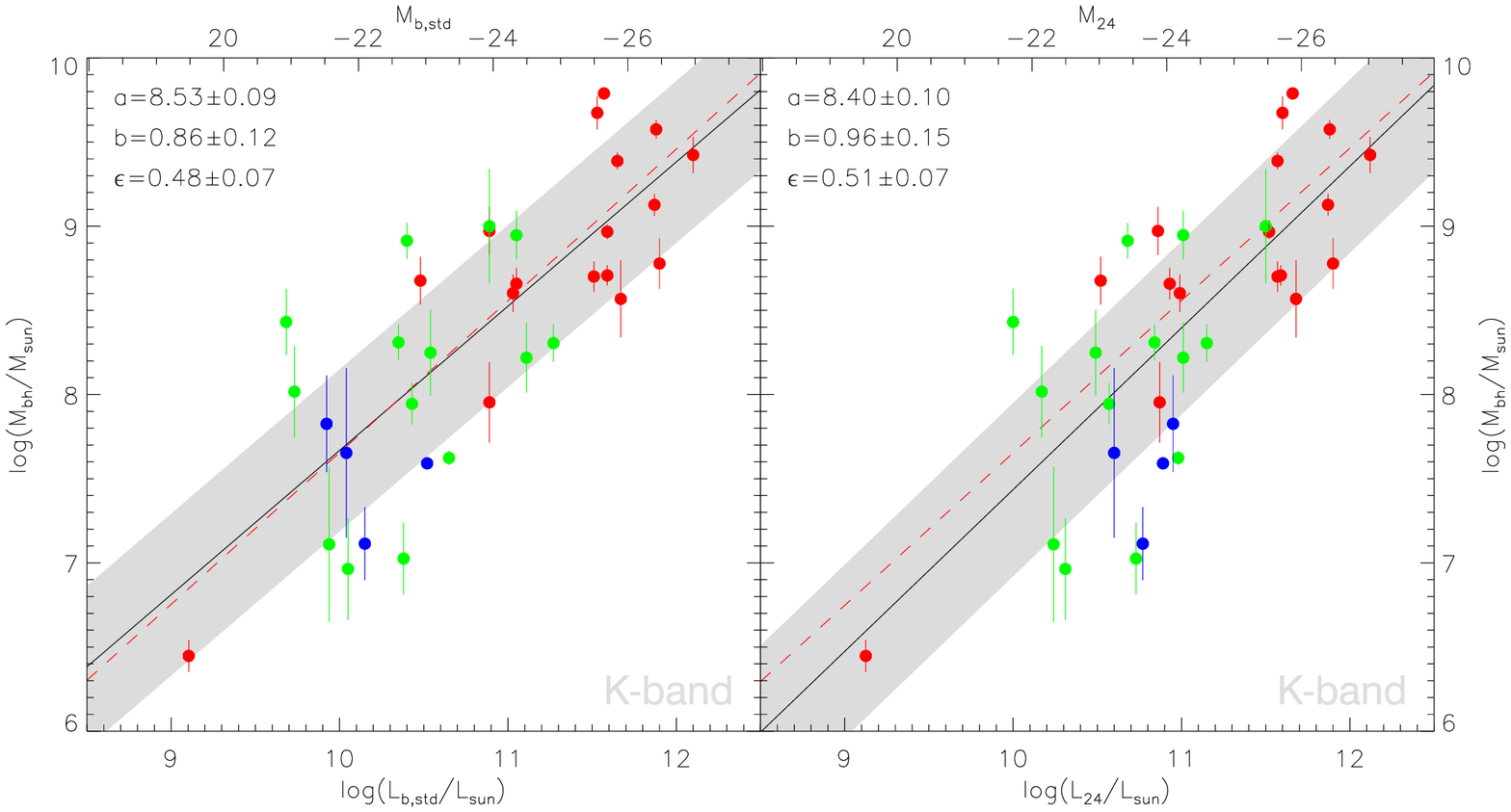}
  \end{center}
  \caption{Correlation between $\mbh$ and ``standard'' bulge  ($\lbstd$ left panel) and non-parametric isophotal ($\liso$, right panel) luminosity. The filled circles correspond to elliptical (red), S0 (green) and spiral (blue) galaxies. Vertical solid bars indicate the $1\sigma$-uncertainties in $\mbh$. The black solid lines are the corresponding best-fit linear relations of the form $\log \mbh=a+b(\log L - 11)$. The shaded area has a width of $2\epsbh$ in the direction of $\mbh$. As expected, the best-fit slope $b$ is steeper for total luminosities. Crucially, the intrinsic scatter $\epsilon$ of both relations is the same. The red dashed lines are the relations fitted to the sample of elliptical galaxies only (see \S\ref{subsec:mainres} and Table \ref{tab:orig_full}).}
  \label{fig:mainfig}
\end{figure*} 

Comparing how $\mbh$ correlates with $\lbul$ and $\ltot$ in the $K$-band, we find three significant results:\\
\begin{enumerate}
 \item bulge and total luminosities lead to similarly tight correlations with comparable intrinsic scatter;
 \item the log-slope of the $\mbh-\lbul$ relation is significantly less than unity, while the $\mbh-\ltot$ relation is consistent with being linear; and
 \item the characterization of the $\mbh-\ltot$ relation is robust against the choice of method used in deriving  $\ltot$. This is not the case for the $\mbh-\lbul$ relation.
\end{enumerate}

\begin{figure*}
  \begin{center}
    \includegraphics[width=18cm]{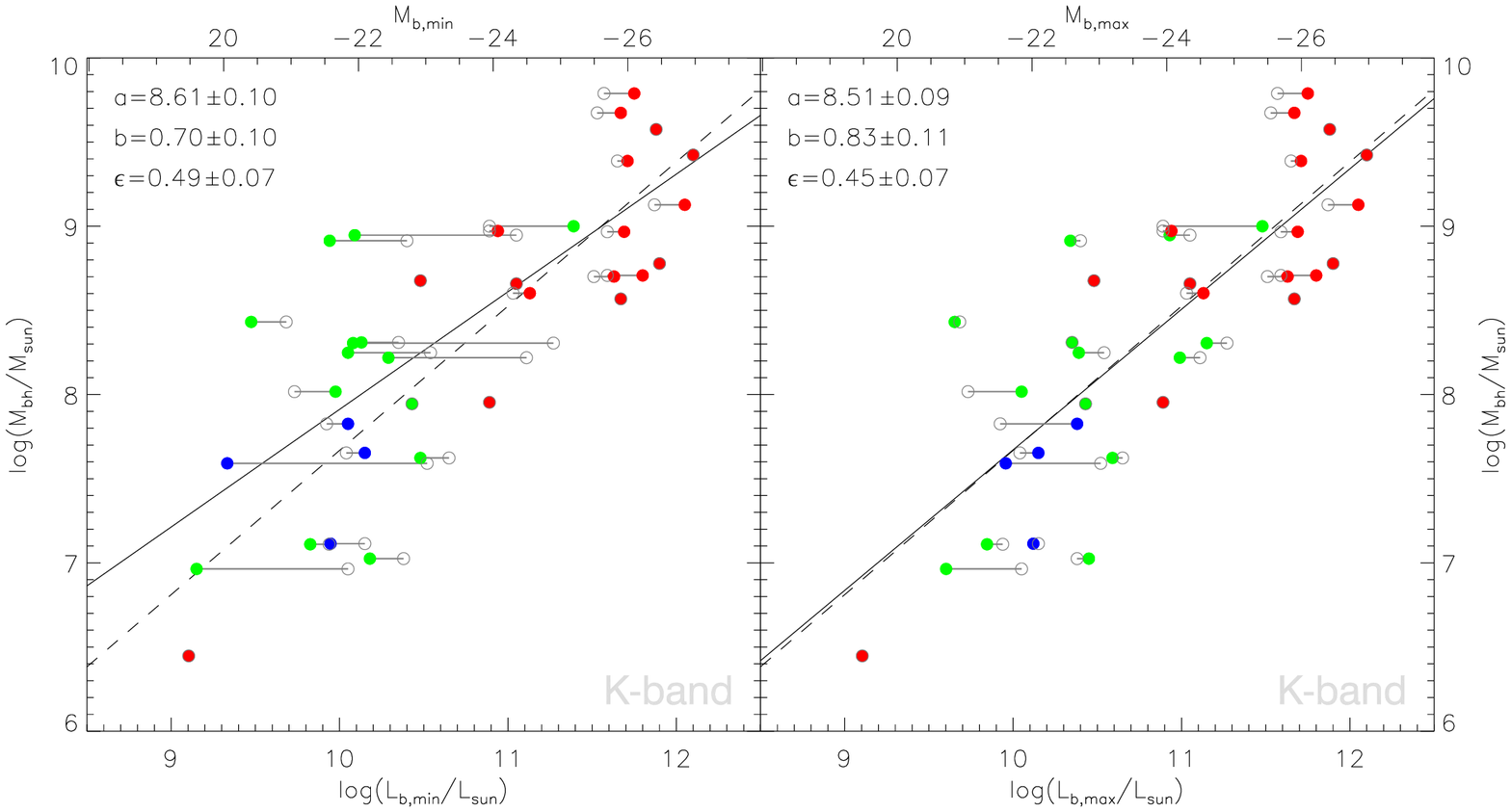}
  \end{center}
  \caption{Correlations of $\mbh$ with improved models' minimal ($\lbmin$, left panel) and maximal ($\lbmax$, right panel) bulge luminosities. Filled circles, error bars and black solid lines are as defined in Figure \ref{fig:mainfig}. Overplotted in gray are the standard bulge values (open circles), while the dashed line is the corresponding $\mbh-\lbstd$-relation. Error bars in $\mbh$ have been omitted for clarity, but are the same as in Fig. \ref{fig:mainfig}. Note that, somewhat unexpectedly, the $\mbh-\lbmax$-relation is consistent with the $\mbh-\lbstd$ relation. The figure also illustrates the changes in bulge luminosity as a consequence of applying a detailed decomposition. Notably, some $\lbmin$ are \textit{larger} than $\lbstd$, while some $\lbmax$ are \textit{smaller} than $\lbstd$. One also sees the effect of cores in elliptical galaxies (red circles), which are masked when obtaining $\lbmin$ and $\lbmax$.}
  \label{fig:mlbul}
\end{figure*}

These results hold regardless of the method used to derive the luminosity (standard/improved models or curve-of-growth analysis). Points (i) is highlighted in Figure \ref{fig:mainfig} using $\lbstd$ and $\liso$, as these represent the simplest available estimates of bulge and total luminosity, respectively. Figure \ref{fig:mlbul} illustrates point (iii), i.e. the dependency of the $\mbh-\lbul$ relation on the details of the photometric decomposition, which causes large (and not easily quantifiable) uncertainties in the  bulge photometry. Finally, confirming (i) and demonstrating result (ii), Figure \ref{fig:adopted} displays our adopted (preferred) $\mbh-\lsph$ and $\mbh-\ltimp$ relationships based from our improved decompositions.

The intrinsic scatter does not seem to depend on which choice of luminosity is adopted in defining the relations. In particular, the scatter found using bulge luminosities derived from the ``standard" model, $\eps\bstd=0.48^{+0.10}_{-0.04}$, is virtually identical to the one obtained when using bulge luminosities from the ``improved" models, whether they are the minimum (i.e. bulge alone), maximum (i.e. bulge plus additional components except the disk and spiral arms), or "spheroidal" (bulge plus envelope) estimates (see \S\ref{subsec:datadecomp}). This is somewhat surprising, given that the improved models generally do a significantly better job than standard models at reproducing the structure seen in many of our sample galaxies  (see Paper-I). 

If total magnitudes are used, likewise all choices of luminosity lead to $\eps\tot\approx0.5$, again consistent with $\eps\bstd$. In other words, no matter whether bulge or total magnitudes are used, and regardless of the specifics of the decomposition or photometric analysis, \textit{all} $\eps$ agree at the 68\%-confidence level. The same applies when one considers $\epso$, the scatter perpendicular to the ridge line (see Table \ref{tab:epsortcomp}). Both $\epsbh$ and $\epso$ decrease when the fit is restricted to elliptical galaxies, but the difference is not significant beyond the $1\sigma$-confidence level.

While the scatter does not seem to be sensitive to the details of the photometric analysis, the characterization of the relation, in particular its slope, {\it does}, but only when bulge luminosities are used. When using total luminosity, all $\mbh-\ltot$ relations agree closely in slope, $b\tot$, irrespective of the method used to obtain $\ltot$. The slopes of the $\mbh-\lbul$ relations obtained from the improved models, $b\bmin$ and $b\sph$, are smaller than $b\tot$ at the $1\sigma$-confidence level. This is to be expected: while the high mass end is dominated by elliptical (i.e. pure bulge) galaxies, bulges comprise only a fraction of the total luminosity in the (on average less luminous) lenticular and spiral galaxies. $b\bul$ increases when maximum bulge magnitudes or standard decompositions are used, such that $b\bmax$ and $b\bstd$ are consistent with $b\tot$ and $b\ellip$.

Using improved decompositions (Figure \ref{fig:adopted}), the slope of the $\mbh-\ltot$ relation remains virtually unchanged whether all galaxies, or only ellipticals are fitted. Consequently, the slope of the $\mbh-\lbul$ relation ($\lbul=\lsph$), when fitted to the entire sample, is also systematically smaller (at the $\sim1\sigma$ confidence level) than the slope fitted to the sample of ellipticals only, $b\ellip$.  Yet, $b\ellip$ should be treated with caution, since it is very sensitive to a specific galaxy, NGC221 (M32), the only elliptical in our sample with $L\lesssim10^{11}\lsun$ and $\mbh\lesssim10^8\msun$. Therefore, we view the evidence that the slope of the $\mbh-\lbul$ relation depends on Hubble type as tentative at best.

\subsection{The adopted scaling relations}
\label{subsec:adopted}
Our recommendation is to use $\mbh-\lbul$ scaling relations based on ``improved model'' fits to the $K$-band images. The rationale for introducing galaxy image models that improve on the ``standard'' bulge(+disk) configuration are described in Paper-I. In short, they are warranted by considerable and characteristic mismatches between data and standard models, with residuals usually revealing bars, galactic nuclei (AGN or clusters), inner disks and spiral arms. Such components are often easily identified by visually inspecting the science image. They are known to be morphologically and kinematically distinct from  the spheroidal (``hot'') stellar component and should therefore be excluded when characterizing the properties of the latter. Such considerations do not apply to elliptical galaxies, although we note that masking the cores in core-\sersic\ galaxies \citep[e.g.][]{Ferrarese_etal06b} improves the luminosity estimate, eliminating the bias induced by the \sersic\ profile's mismatch in the central region.

As for which estimate of the bulge luminosity to use, our preference is for $\lsph$, the luminosity of the bulge component plus, when present, an additional spheroidal envelope, for the following reasons. $\lbmax$ does not measure the luminosity of the bulge/spheroidal component, but of all components except the disk and spiral arms. It therefore marks an upper limit for $\lbul$ and was introduced primarily to compare results from standard and improved models. $\lbmax$ \textit{would} be the ``proper'' luminosity only \textit{if} one assumed that the extra components (such as bars and stellar nuclei) are not separate entities but are instead part of the bulge. By contrast, $\lbmin$ is the luminosity of the central component with \sersic\ index $n>1$ and higher axis ratio than the disk(s), and therefore corresponds to the commonly adopted photometric definition of ``bulge''. As described in Paper-I, for five of our galaxies this definition probably does not describe the \textit{spheroidal} (``hot'') stellar component entirely. Instead, an ``envelope'' component is required for suitable image modeling and should be added to $\lbmin$ in order to measure the luminosity of the spheroid, $\lsph$ (we note that the physical nature of such envelope is not clear -- it might simply be part of the bulge, or even an artifact introduced by the choice of parametrization for the bulge and the disk, see the discussion in Paper-I). Therefore, interpreting the term ``bulge'' in the sense of ``spheroid'', we adopt the correlation of $\mbh$ with $\lsph$, rather than $\lbmin$, as the best available characterization of $\mbh-\lbul$.

\begin{figure*}
  \begin{center}
    \includegraphics[width=18cm]{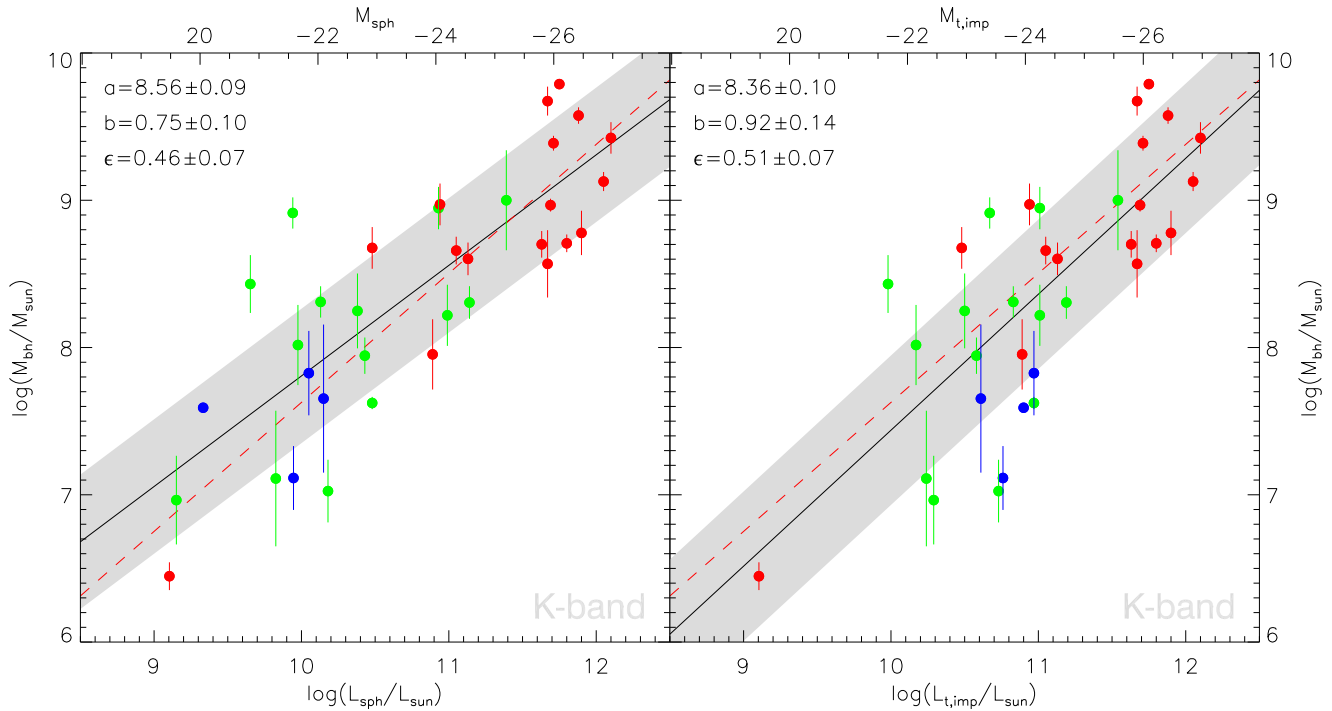}
  \end{center}
  \caption{Our adopted correlations of $\mbh$ with bulge/spheroid ($\lsph$, left panel) and total ($\ltimp$, right panel) luminosity. Filled colored circles, error bars and lines are defined as in Figure \ref{fig:mainfig}. The respective intrinsic scatter is indicated by the shaded area, which has a width of $2\epsbh$ in the direction of $\mbh$. Note the similar scatter of the two relations, as well as the similarity between the slopes of $\mbh-\ltimp$ and the fit to ellipticals only, $\mbh-\limp\ellip$ (red dashed line in both panels).}
  \label{fig:adopted}
\end{figure*}

When using total luminosity, we also prefer values derived from improved models, for the same reasons given above, although we note that correlations with total luminosities from standard models and aperture photometry lead to virtually identical correlation parameters. The $K$-band $\mbh-\lsph$ and $\mbh-\ltimp$-relations are highlighted in Tables \ref{tab:lumdef}, \ref{tab:orig_full} and \ref{tab:rels_alt}, and plotted in Figure \ref{fig:adopted}.

\subsection{Comparison with previous studies}
\label{subsec:rescomp}

\begin{table*}
\centering
\caption{Comparison with previous studies}
\begin{minipage}{17cm}
\renewcommand{\thempfootnote}{\fnsymbol{mpfootnote}}
 \begin{tabular}{cl*{10}cl}
  \toprule
    & ref. & imaging & sample & $N$ & fit & $a$ & $\Delta a$ & $b$ & $\Delta b$ & $\epsbh$ & $\Delta\epsbh\,(+/-)$ & remarks \\   
  \input{table_litcomp}
  \bottomrule
 \end{tabular}
 \end{minipage}
 \tablecomments{$\mbh-\lbul$ relations from previous studies (lines 1-8), from common sub-samples (lines 9-12) and our adopted result (boldface). Column (1) gives the source of the fit results, where MH03: \protect\cite{MH03}; G07: \protect\cite{Graham07}; S11: \protect\cite{Sani_etal11}; V12: \protect\cite{V12}; and ``-``: this work. The imaging data used to derive the luminosities are listed in column (2). Column (3) indicates if a subsample was used in acquiring the fit (``all'': the full sample of $\mbh$ considered in the respective publication; ``rel'': subsample of reliable $\mbh$ estimates as indicated by the authors; ``res'': sample restricted based on criteria not related to the reliability of $\mbh$; ``$\cap$'': intersection of our sample of reliable $\mbh$ and MH03's full sample), and column (4) the resulting sample size. Column (5) indicates the method used in acquiring the parameter estimates, where the entries are ``AB96``: BCES algorithm of \protect\cite{AB96}; ``T02``: modified FITEXY routine introduced by \protect\cite{T02}, with errors on $\epsbh$ obtained by setting $\chi^2_\nu=1\pm\sqrt{2/N}$; and ``like``: maximum-likelihood / Bayesian analysis, whereas we set uncertainties $\Delta L=0$ as discussed in \S\ref{subsec:error_treat}. Columns (6)-(11) present the correlation parameters and $1\sigma$-uncertainties, with values from the literature converted to our convention, $\log\mbh=a+b\log(L/10^{11}\lsun)$, if necessary. Column (12) supplements information on the data set on which the fit was based. If not stated otherwise, $\mbh$ and $L$ are the values used by the source given in column (1). The bulge luminosities from this work (last three lines) are our adopted $\lsph$.}
 \label{tab:litcomp}
\end{table*}

In Table \ref{tab:litcomp}, we compare our results (bottom line) with those from previous studies that investigated the $K$-band $\mbh-\lbul$ scaling relation (lines 1-8). We also include in the comparison the relation of \cite{Sani_etal11} [S11 hereafter], which was derived at $3.6\mu\mathrm{m}$.

Our zero-point ($a=8.56\pm0.09$) is consistent with the value measured by V12, but is higher than that of MH03  and, in particular, \cite{Graham07} [G07 hereafter], who re-fitted the data from MH03 after revising a number of $\mbh$ and distance values, and removing five galaxies from MH03's sample of reliable $\mbh$ measurements. The results of MH03 and G07 were both based on \twomass\ data (G07 relied on the 2D fits of MH03), which is inferior, both in spatial resolution and depth, to our data or the \ukidss\ data used by V12. In Paper-I, we show that using data with low S/N or spatial resolution typically leads to overestimate the bulge luminosities, and speculate that this may explain the lower zero-point resulting from \twomass\ data.

Our log-slope, $b\sph=0.75\pm0.10$, is shallower than found in all previous studies. The difference is above the $1\sigma$-level, and again the smallest difference is found with respect to the V12 estimate. The comparison suggests a general trend towards shallower slope with increasing data quality and decomposition complexity. G07 obtains a shallower slope than MH03, presumably in part a result of separating\footnote{by employing other (optical) data} the disk from the bulge in three galaxies to which MH03 applied a single \sersic\ profile. In contrast to MH03 and G07, the resolution and depth of the data allowed S11 and V12 to also model nuclei and bars, and they find a (slightly) shallower slope than G07. V12 also accounted for cores, and their result yields $b<1$ at the $1\sigma$-confidence level. Our slope is shallower still, possibly due to further improved bulge photometry, which typically reduces the luminosity estimate at the lower-mass end (predominantly late Hubble type) and increases it for some of the ellipticals.

We note that the choice of fitting method also plays a role, as expected \citep[e.g.,][]{NFD06}: When fitting the \textit{same} data ($L_K$ \textit{and} $\mbh$) as used by MH03 (their Table 1) with our likelihood method\footnote{for consistency, we choose not to incorporate uncertainties in $L$, see \S\ref{subsec:error_treat}} (lines 2 and 4 of Table \ref{tab:litcomp}), the resulting slopes are shallower than the relation reported by MH03 \citep[bisector estimator of][lines 1 and 3]{AB96}, although they are still significantly steeper than the relations derived in this paper.

As for the scatter, we again are in agreement with the result V12, deriving a larger $\eps$ than either MH03 (for their "group 1" sample) or G07, despite using improved data and decompositions. The scatter derived by S11 using $3.6\mikron$ data also agrees more closely with ours, although it is important to keep in mind that both S11 and V12 use samples and sample selection criteria that are different from one another and from our study. S11 exclude pseudobulges identified based on the bulge's \sersic\ index, which may be highly uncertain and subject to the modeling complexity (see Paper-I). V12 exclude barred galaxies, M87 (due to concerns about the background subtraction) and an outlier, NGC4342. Irrespective of the suitability of the applied criteria to identify pseudobulges / barred galaxies, and the astrophysical justification to exclude such objects from the analysis, these sample restrictions are expected to decrease the scatter, since by construction they do not apply to elliptical galaxies, and (presumably) only rarely to early-type lenticulars. That is, they tend to bias the sample towards early-type and luminous galaxies, for which numerous studies (including ours) find a slightly decreased scatter. Indeed, V12 find a marginally higher scatter ($0.52^{+0.10}_{-0.06}$ instead of $0.40^{+0.09}_{-0.06}$) when analyzing their full sample of reliable $\mbh$ (line 5 in Table \ref{tab:litcomp}), higher than the scatter of our adopted $\mbh-\lsph$ relation ($0.42^{+0.08}_{-0.04}$).

To investigate how the choice of decomposition affects the slope, normalization, and scatter of the relation, we extract a sample in common between ours and  MH03's, and derive scaling relations for this sample using both our data and MH03's (lines 9-12 in Table \ref{tab:litcomp}). Using MH03's black hole masses and luminosities, but restricting the sample to the common objects (from 37 to 28 galaxies, compare lines 2 and 9) leaves the slope $b$ unchanged, but increases the normalization $a$ and decreases the scatter $\eps$ at the 1$\sigma$ level (compare line 9 to line 2 in table \ref{tab:litcomp}). We then replace the $\mbh$ that MH03 used with our updated values (line 10), and find little effect on $b$ and $\eps$, but a significantly increased zero-point. If we use the same $\mbh$ as MH03, but replace the MH03 luminosities with our improved $\lsph$ (line 11), the zero point also increases, but not quite as much. At the same time, $b$ and $\eps$ decrease by $0.11$ and $0.04\dex$, comparable to the change we observed when replacing standard bulge luminosities ($\lbstd$) with $\lsph$ while fitting our (full) sample. Using both improved $\lsph$ and updated $\mbh$ (line 12, compared to line 9) combines the above effects (increased $a$, decreased $b$, slightly decreased $\eps$) and leads to a best-fit relation that is in close agreement with our adopted relation (our full sample, line 13). Therefore, the above analysis confirms that accounting for components beyond the bulge and the disk (as done in our study but not in MH03) is the dominant driver towards a shallower slope, while the updated $\mbh$ estimates are the main (but not sole) cause for the higher zero-point of our relation relative to that of MH03. Meanwhile, the intrinsic scatter appears to be primarily effected by sample selection and decreases only slightly when using improved bulge luminosity estimates.

\section{DISCUSSION}
\label{sec:discussion}

\subsection{The theorist's perspective}
\label{subsec:theo_q}
Our analysis suggests that $\mbh$ correlates as tightly with $\ltot$ as it does with $\lbul$. This is evidenced by the fact that the intrinsic scatter (whether it is measured along the $\mbh$ axis or orthogonal to the relation) for the two relations is comparable: for instance, along the $\mbh$ axis, $\eps\bul=0.46^{+0.09}_{-0.04}$ while $\eps\tot=0.51^{+0.10}_{-0.04}$ (lines 4 and 7 of Table \ref{tab:orig_full}). This result challenges the view that \smbh s do not scale as tightly (or at all) with the total galaxy luminosity, as first argued by \cite{KormendyGebhardt01} based on $B-$band data from a non-identified source. The use of $B-$band luminosity, however, is not-ideal to address this issue, at least if it is assumed that mass, rather than luminosity, might be the primary driver for the relations. Subsequent studies have, therefore, correctly focused on luminosities derived at longer wavelengths. 
\cite{Beifiori_etal12}, using $i-$band data, state that ``bulge luminosity is a better tracer of the mass of the \smbh s than total light of the galaxy". However, this claim seems to be contradicted by their Table 5, which shows that the \textit{intrinsic} scatter in the relation using bulge magnitudes ($0.58 \pm 0.11$ for their sample ''B'') is, at face value, larger than the scatter in the relation using total magnitudes ($0.52 \pm 0.11$). The \textit{total} scatter of both relations is consistent in all samples used by \cite{Beifiori_etal12}, and similarly when luminosities are replaced by the respective stellar or dynamical mass.

The connection between $\mbh$ and total stellar mass, $M_{\star,\mathrm{tot}}$ (and implicitly its surrogate, $\ltot$), has been the subject of studies investigating BH scaling relations for active galaxies and their evolution with redshift \citep[e.g.][]{Kim_etal08a,Merloni_etal10,Bennert_etal11,Cisternas_etal11}, and some have suggested that it might represent a comparable or even tighter correlation than $\mbh-\Mbul$ \citep{Peng07,Jahnke_etal09,Bennert+10}. Our result provides a local baseline for these studies, with the caveat that it is based on quiescent galaxies and different $\mbh$ measurement methods. It further seems to support the idea that \smbh\ growth may be linked with the overall potential of the host \citep[as in][]{Ferrarese02,VolNatGul11b}, as traced by its total NIR luminosity (stellar mass). This is also in line with the evidence that, at fixed radiative efficiency, the integrated  accretion history of \smbh s, as traced by active galactic nuclei, parallels the full star formation history of \smbh\ host galaxies \citep[e.g.,][]{ShankarWeinbergMiraldaEscude11}. An $\mbh-M_{\star,\mathrm{tot}}$ relation with scatter similar to that of $\mbh-M_{\star,\mathrm{bul}}$ is also predicted by the merger-driven models of \cite{JahnkeMaccio11}. We note however that, if repeated sequences of galaxy mergers have characterized the growth of spheroids and disks, and have been strictly accompanied by mergers of their central \smbh s, the ensuing relations should become tighter for elliptical (and more massive) galaxies, while we observe only a modest decrease. However, firm conclusions on this point must await larger samples of both early and late type galaxies, spanning a wider mass range than the present sample. 

Our results for the correlation slope and offset also impact our current understanding of the connection between \smbh\ and host galaxy. The slope we obtain for (the logarithm of) the $\mbh-\lbul$ relation, $b\bul = 0.75 \pm 0.10$, excludes a direct proportionality between $\mbh$ and bulge stellar mass, $M_{\star,\mathrm{bul}}$, with high confidence. Assuming that the $K$-band stellar mass-to-light ratio, $M_\star/L_K$, is on average positively correlated with $L_K$, the log-slope of the $\mbh-M_{\star,\mathrm{bul}}$ relation is even lower than $b\bul$, which is already $<1$ with $\gtrsim99\%$ confidence according to our study. Combining our $\mbh-\lbul$ relation with, for example, $\log M_\star\propto(1.12\pm0.02)\times\log L_K$ \citep[][]{Zhu_etal2010}, a simple estimate yields $\log\mbh\propto(0.67\pm0.11)\log M_{\star,\mathrm{bul}}$. That is, the $\mbh/M_{\star,\mathrm{bul}}$-ratio \textit{decreases} with host bulge/spheroid mass. This is in agreement with \cite{Sani_etal11}, who found $b_{\star,\mathrm{bul}}=0.79\pm0.09$, but contrasts with the results of \cite{HR04}, who measure bulge masses directly from dynamical modeling and find a log-slope significantly greater than unity. Using the same $L_K \rightarrow M_\star$ conversion as above, we find that the $\mbh-M_{\star,\mathrm{tot}}$ relation also has a log-slope below unity, albeit with lower significance: $b_{\star,\mathrm{tot}}\approx0.82\pm0.15$. This seems to imply that \smbh s build up their mass at a lower relative rate than the stars (bulge), at least in the range considered here ($M_\star\gtrsim10^{9}\,M_\odot$), or that the BH seeds were relatively smaller in the presently most massive galaxies.

\subsection{The observer's perspective}
\label{subsec:obs_q}

Which quantity is a more efficient and reliable $\mbh$ indicator? According to the present work, $\ltot$ should be preferred to $\lbul$, for very pragmatic reasons. The detailed decompositions and careful case-by-case analysis needed to separate the bulge light is all but unfeasible in most cases in which one wishes to estimate $\mbh$, especially for large sample of galaxies, or galaxies at higher redshifts. For such programs, anything beyond an automated decomposition into bulge(+disk) is unrealistic. However, the $\epsbh$ expected when $\mbh$ is measured using this ``standard'' decomposition is virtually identical to the scatter  in the $\mbh-\ltot$ relation, which negates the advantage of performing a decomposition in the first place. In fact, even if a more detailed decompositions were possible, the predictive power of the measured $\lsph$ would improve only slightly over that offered by $\lbstd$ ($\eps_\mathrm{sph}=0.46$ instead of $\eps_\mathrm{b,std}=0.48$, compared to $0.51\dex$ scatter in $\mbh-\ltot$). Moreover, the slope of the relations involving $\lbul$ depends on the details of the decomposition, which by itself is dictated by the quality of the data. This is likely to introduce biases in the results. An additional argument against the use of the $\mbh-\lbul$ relation to estimate \smbh\ masses is the possible dependence of its parameters on the mix of morphological types (ellipticals versus all galaxies). If such dependence is confirmed, studies of \smbh\ demographics based on the $\mbh-\lbul$ relation may suffer from enhanced systematic error, since the local sample of $\mbh$ is not representative of the actual distribution of Hubble types in the universe. 

By contrast, the scaling relation with $\ltot$ appears to be less dependent on Hubble type, is more robust with respect to the employed photometric/decomposition method (see Paper-I), and less dependent on the available image quality (depth, resolution). Furthermore,  $\ltot$ can be measured much more efficiently: for example, it can be obtained non-parametrically using a curve-of-growth approach ($\liso$), at least if, as in our study, the background can be estimated reliably and potential interlopers / nearby objects can be masked. In many cases, non-parametric estimates of total luminosity may be more accurate than parametric measurement based on profile fits (e.g. in galaxies with bright AGNs, embedded disks, strong bars or spiral arms).

Apart from reliability and practical considerations, both our $\mbh-\ltot$ and updated $\mbh-\lbul$ relations  have direct consequences on estimating the (local) \smbh\ mass function (Shankar et al., in prep.). As reported, for example, by \cite{Tundo_etal07} and \cite{Lauer_etal07a}, combining the luminosity function of bulges with the  $\mbh-\lbul$ relation predicts a significantly larger \smbh\ mass density than is obtained by combining the velocity dispersion function with the $\mbh-\sigma_\star$ relation, especially at high \smbh\ masses. Both the lower slope $(b)\bul$ of our updated relation, and use of the $\mbh-\ltot$ relation (which has a slope comparable to the one previously measured for $\mbh-\lbul$), may reduce the discrepancy in this regime\footnote{see also, for example, \cite{G09} regarding the strong impact of the intrinsic scatter}.

\subsection{Systematic uncertainties and caveats}
\label{subsec:sys_uncertain}

As mentioned in \S\ref{subsec:mainres} and \S\ref{subsec:obs_q}, the slope of the $\mbh-\lbul$ relation tends to be biased too high  unless a detailed galaxy decomposition is performed. We stress that this bias does not affect only  spiral galaxies. Spiral arms are detected only in 4 of our 35 targets, but in spite of this, we  still find a significantly different slope by using $\lbstd$ instead of $\lsph$ or $\lbmin$. It is difficult to assess whether or to what degree our decompositions included all relevant components, and thus whether a bias in slope  remains. It also must be kept in mind that the surface brightness profiles of bright early type galaxies deviate from a \sersic\ law in the innermost few arcseconds; forcing a 2D-\sersic\ profile fit without masking the core in these galaxies, affects the slope of both relations (with $\lbul$ and $\ltot$), although not to as large a degree as the details of the decomposition for galaxies with multiple morphological components (cf. $b\timp$ with $b\tstd$ in Table \ref{tab:orig_full}). Overall, the slope of the relation between $\mbh$ and total (or isophotal) magnitudes is less affected by systematic errors than the slope of the relation involving bulge (or spheroid) magnitudes.

In addition, even when a detailed photometric decomposition is available, it is not always straightforward to decide which components should be included in the ``bulge" luminosity. A specific example are the ``envelopes" fitted in the majority of lenticular galaxies. In \S\ref{subsec:adopted} and Paper-I we argued that envelopes should be included in the spheroidal luminosity, and therefore $\lbul=\lsph$ instead of $\lbul=\lbmin$. However, this choice can be debated. The envelope might represent a thick disk in some cases, especially when its axis ratio is intermediate between the bulge and disk. It might even be a spurious component introduced by our requirements that the profiles of bulge and disk must each be represented by a single component with \sersic\ and exponential law, respectively. Not counting the envelope as part the bulge leads to lower bulge luminosities preferentially for galaxies at the fainter half of the sample, and therefore lowers the slope of the $\mbh-\lbul$ relation (lines 2 and 4 of Table \ref{tab:orig_full}).

The above considerations also lead us to emphasize that all of our bulge measurements rely on the universal application of the \sersic\ profile. In bulge dominated galaxies, this assumption may be tested, and the \sersic\ parameters may be derived reliably. However, in many galaxies, the bulge is not the dominant component. In this case, not only the assumption that the profile follows a \sersic\ law must be regarded with caution due to the significant overlap with the disk (and additional components if present), but the best fit parameters are often degenerate between components: for this very reason, for instance, bulge+disk decompositions are almost always performed under the assumption that the disks are exponential, since this lifts some of the degeneracy. The end result is that bulge magnitudes are likely affected by potentially serious, and not easily quantifiable random and systematic errors.  

We now turn our attention to the intrinsic scatter of the $\mbh-\lbul$ relation. Given the uncertainties associated with the bulge luminosities, it is fair to ask whether the errors on the latter could have been underestimated: if this is the case, then one might expect the intrinsic scatter in the $\mbh-\lbul$ relation to be lower than reported in this paper. In fact, the effect is negligible. First, going from a bulge+disk to improved decompositions lowers the scatter by only $0.02\dex$, and further improvements from an even more complex photometric analysis are most likely minimal. Furthermore, adopting a coarse estimate of the error on the luminosities lowers the intrinsic scatter by only $0.01\dex$ (see Appendix \ref{subsec:error_treat}). 

Changes in the sample or revised $\mbh$ are likely to have a larger impact on the relation than further improvements in the photometry. In particular, all of our results depend on the accuracy of the available $\mbh$ and their uncertainties. Systematic errors on $\mbh$ may originate from neglecting radial variations of the stellar mass-to-light ratio \citep[see discussion in][]{G09}, from an incomplete orbit library used in the dynamical model \citep[][]{ShenGebhardt10,SchulzeGebhardt11}, or from the assumption of axisymmetry in the case of triaxial galaxies \citep[][]{vdBosch_deZeeuw10}. In the case of $\mbh$ measurements based on gas dynamics, systematic errors might arise from neglecting turbulence, non-gravitational forces, or asymmetries in the gas distribution.  Currently available measurements of $\mbh$ differ in the modeling technique and the statistical treatment of the data: homogeneous modeling of the data, including comprehensive statistical analysis of the uncertainties, is needed to further refine the $\mbh-L$ scaling relations.

\subsection{Pseudobulges}
\label{subsec:pseudobulges}

Whether or not \smbh s in ``pseudo-''bulges follow the same scaling relations as those defined by ``classical" bulges has received significant attention in the literature \citep{Nowak_etal10,K11,Graham12,Benitez_etal13}. Pseudobulges by definition do not occur in elliptical galaxies, and given the current mix of galaxies with secure measurement of $\mbh$,  they are found preferentially at the low-mass end of \smbh\ scaling relations. If \smbh s in pseudo-bulges do not behave as those in classical bulges, including them would necessarily lead to relations with a biased slope, normalization and scatter.

Pseudobulges are  identified based on morphology: they have low \sersic\ index ($n < 2$), and are associated with distinct morphological features, including nuclear bars, spiral structures, dust, and flattening similar to the disk \citep{K11}. As discussed in Paper-I, however, identifying pseudobulges based on these criteria is fraught with uncertainty. Based on our photometry, 4 out of 35 galaxies have a bulge best described by $n < 2$, although this is critically dependent on the details of the decomposition. In particular, if a standard bulge+disk decomposition were to be performed, {\it none} of these four galaxies would be described by an $n < 2$ bulge, while {\it three} different galaxies {\it would}, making the classification based on morphological structure alone problematic. Furthermore, in studies of BH demographics the population of galaxies with pseudobulges would need to be identified in an automated manner based on bulge+disk decomposition, the usefulness of which, as mentioned above, is questionable.  

In the context of \smbh-galaxy co-evolution models, it is unclear whether and how pseudobulges follow a separate relation than the one defined by classical bulges. This is not surprising, considering the range of sample and sub-sample selection as well as data quality and image analysis methods, which we have shown in this work to impact the derived structural parameters of (pseudo-)bulges and hence the identification of pseudobulges to begin with. We also caution that pseudobulges may co-exist with classical bulges \citep[see e.g.][]{Nowak_etal10}, a result that seems to be confirmed by our work: an inner disk component (which many might call a ``pseudobulge") is present in 6 of 18 galaxies with a (classical) bulge and large-scale exponential disk. We conclude that until we have clearly defined and consistently measurable pseudobulge criteria that are shown to select a population with a physically distinct origin and/or evolution, fitting \smbh\ scaling relation to galaxy subsamples based on the presence of a pseudobulge is not warranted.

\section{Summary and conclusions}
\label{sec:summary}

We have revisited the $\mbh-\lbul$ relation using a sample of 35 nearby galaxies of all Hubble types, for which we have obtained new, deep $K-$band photometry using WIRCam at CFHT. Our analysis benefits from a homogenized database of black hole mass estimates, as well as from the depth and spatial resolution of the near-infrared data, which significantly exceed those of the \twomass\ or \ukidss\ images used in previous investigations. A detailed morphological decomposition was performed for each galaxy using \gfthree, leading to the identification of multiple  components (nuclei, bars, small-scale disks and envelopes, in addition to the canonical bulge and large-scale disk) in a significant fraction of objects (Paper-I).

We find that the zero-point and (logarithmic) slope of the $\mbh-\lbul$ relation differ significantly from those published in \cite{MH03} and \cite{Graham07} (both based on \twomass\ data), but are consistent with the values measured in the recent study by \cite{V12} (based on \ukidss\ data). In particular, we find at the 99\% confidence level that \smbh\ masses increase more slowly than the $K$-band bulge luminosity. If we assume that $M_\star/L_K$ is constant or increases with galaxy luminosity, this implies that the correlation between black hole mass and bulge mass, $\mbh-M_{\star,\mathrm{bul}}$, also has a slope below unity: the most massive galaxies seem to have proportionally lower black hole masses per given stellar (bulge) mass.

We find a maximum-likelihood scatter in the relation of $0.48\dex$ (in log $\mbh$) for a straightforward bulge(+disk) estimate of $\lbul$, marginally decreasing to $0.46\dex$ if more sophisticated approaches to extract the bulge luminosity are used. This scatter is larger than reported by MH03 ($0.3\dex$), in spite of the superior photometric data and analysis, and the improved $\mbh$ adopted in our work. However, our estimate of the scatter is again consistent with the $K$-band analysis of \cite{V12}, as well as with recent results in the optical \citep[][]{G09} and at $3.6\mikron$ \citep[][]{Sani_etal11}. We stress that our results are based on a sample that is unbiased with respect to the morphological classification of the host galaxy: our only criterium for inclusion of a galaxy is the availability of a precise measurement of the black hole mass. By contrast, the studies mentioned above have often culled the sample by removing outliers or specific morphological classes (most notably ``pseudobulges''), which likely lowers the scatter in the relation. Preference for our approach originates, among others, in the envisaged use of the relation for \smbh\ demographic studies.

We have also investigated the relation between Supermassive Black Hole masses and {\it total} galaxy luminosity, $\mbh-\ltot$, finding that within the uncertainty, $\mbh$ appears to be linearly proportional to $\ltot$ (log-slope is unity). The scatter of the $\mbh-\ltot$ relation is consistent with that of the $\mbh-\lbul$ relation, regardless of the details of the photometric decomposition. This result is at variance with previous studies, and might imply that total, rather than bulge, luminosity (mass) is the fundamental driver in the co-evolution of galaxies and black holes. This notwithstanding, we find that while the characterization of the $\mbh-\lbul$ relation depends on the details of the photometric analysis, the $\mbh-\ltot$ relation does not, i.e. it can be characterized more robustly. Combined with the uncertainties and pitfalls inherent to a parametric decomposition of a galaxies into separate morphological components (the results of which depend on the quality -- depth and spatial resolution -- of the data), we 
suggest that the $\mbh-\ltot$ relation should be preferred in studies of \smbh\ demographics, and should be adopted as a constraint for theoretical investigations of \smbh-galaxy co-evolution.

\section*{Acknowledgments}
\label{sec:acknowledgments}

RL thanks Arjen van der Wel and Remco van den Bosch for helpful comments. FS acknowledges support from a Marie Curie grant. This research has made use of the NASA/IPAC Extragalactic Database (NED) which is operated by the Jet Propulsion Laboratory, California Institute of Technology, under contract with the National Aeronautics and Space Administration. 


\bibliographystyle{apj}
\bibliography{Mbh-Lbt,Mbh}


\appendix

\section{Fitting method}
\label{app:fitmethod}

\subsection{Choice of fit method}
\label{subsec:fitmethod}

We initially fit our scaling relations with the modified \fitexy\ routine \citep[see][T02 hereafter]{T02}. \fitexy\ \citep[][]{NumericalRecipies2_Fortran} is a least-squares minimization algorithm that accounts for Gaussian uncertainties in both coordinates and gives estimates of the uncertainties in the derived linear relation parameters $(a,b)$. We use the implementation available in the IDL Astronomy User's Library\footnote{http://idlastro.gsfc.nasa.gov/ftp/pro/math/fitexy.pro}. The routine was modified to account for the intrinsic scatter $\eps$ by adding it in quadrature to the data's y-errors, and finding its "true" value by varying $\eps$ until $\chi_\nu^2=1$.

Broadly following \cite{G09} [G09 hereafter], we also fit all relations via the maximum-likelihood method, for the following reasons:
\begin{itemize}
 \item[(1)] The modified \fitexy\ algorithm does not provide a confidence interval for $\eps$. Although this can be estimated by applying Monte-Carlo resampling (which we did using $n=10000$ samples), the bootstrap estimate may be biased (see next point).
 \item[(2)] For each bootstrap sample, \fitexy\ produces values for $\eps$ that are systematically smaller (by $\sim5-10\%$ on average) than the value obtained by fitting the full sample, possibly because the sample size is too small to produce robust results by bootstrapping. Based on this, one might also question the validity of the bootstrapped confidence limits on $\eps$, as well as the appropriateness of the model \citep[][]{HoggBovyLang10}. 
 \item[(3)] $\eps$ is not a parameter of the model underlying the modified \fitexy\ optimization, but is chosen a priori to give a particular value of $\chi^2$. We are therefore concerned that this way of accounting for $\eps$ may bias the results.
\end{itemize}

The normalization and slope of the relation, $(a,b)$, as well as their uncertainties, derived from the likelihood-method agree closely with those derived from  \fitexy\ (see Table \ref{tab:rels_alt}), which corroborates the finding of G09. Yet, we prefer and adopt the maximum-likelihood method for our quoted results (Table \ref{tab:orig_full}), because it
\begin{itemize}
\item[a)] assumes a generative model which explicitly and naturally includes the intrinsic scatter as one of the model parameters, 
\item[b)] easily accommodates for modifications or generalizations, such as non-Gaussian or correlated data uncertainties, and 
\item[c)] provides complete statistical information in form of the full (3-dimensional) posterior probability distribution, and consequently also straightforward uncertainty estimates of all parameters, including correlation.
\end{itemize}

\subsection{Maximum-likelihood method}

Computation of the likelihood proceeds in a way similar to the method presented in G09, and partially draws from concepts outlined in \cite{calj12}.

The likelihood function, $\mathcal{L}(\{x\},\{y\}\,|\,\mathcal{M}_\theta)$, gives the probability of the measured $N$ data points, $\{x\}=\{x_i\}_{i=1}^N$ and $\{y\}=\{y_i\}_{i=1}^N$, under the assumption that the model $\mathcal{M}$ parametrized by $\theta$ generated the data. In our case, $(x,y)=\left(\log(L/10^{11}\lsun),\log(\mbh/\msun)\right)$, and $\mathcal{M}=\mathcal{M}_{ab\eps}$ is a linear relation between $x$ and $y$, with offset $a$ and slope $b$. The third parameter, $\eps$, characterizes the intrinsic scatter, which we assume to follow a Gaussian\footnote{which is an appropriate distribution as shown by G09} with standard deviation $\eps$ in the $y$-direction. Then, using the the notation
\begin{equation}
  G_\sigma(x)=\frac{1}{\sigma\sqrt{2\pi}} e^{-x^2/2\sigma^2}~,\label{eqn:gaussian}
\end{equation}
our model is the probability distribution of $y$ given $x$ (and the model parameters):
\begin{equation}
 \mathcal{M}_{ab\eps}(y\,|\,x)=G_\eps(y-a-bx)~.\label{eqn:mod_abe}
\end{equation}
This model is not the only possibility to characterize a linear relation with Gaussian intrinsic scatter: $\eps$ could as well be assumed to reside in the $x$-coordinate (luminosity). The resulting fit would differ, as discussed in \cite{NFD06}. Our choice concurs with most recent studies of the $\mbh-\lbul$ relation, and corresponds to the physical interpretation that the galaxy property ($x=\log L$) determines the black hole mass ($y$) via some partially stochastic process. Exploration of the inverse relation, as well as models equivalent to orthogonal least-squares and bisector fits, is beyond the scope of this paper.

Due to measurement uncertainty, which was so far implicit, every datum $(x_i,y_i)$ represents a probability distribution, $P_i(x_i,y_i\,|\,x',y')$, where $(x',y')$ are the true (unknown) values that resulted in the measurements $x_i$ and $y_i$ . In order to obtain the likelihood $\ell_i$ of $(x_i,y_i)$ given the model (\ref{eqn:mod_abe}), we therefore need to integrate over all possible true values:
\begin{equation}
 \ell_i = \int \mathcal{M}_{ab\eps}(y'\,|\,x')\,P(x') P_i(x_i,y_i\,|\,x',y')\, \dd x' \dd y'~, \label{eqn:like_abe}
\end{equation}
where $P(x')$, the probability distribution of (true) $x$-values, has been included such that $\mathcal{M}(y|x)\cdot P(x)$ is a bivariate distribution in $(x,y)$ (see below). If the individual data points and errors are independent, the likelihood of all data is the product of the individual likelihoods:
\begin{equation}
 \mathcal{L}=\mathcal{L}(\{x\},\{y\}\,|\,a,b,\eps)=\prod_{i=1}^N \ell_i~.
\end{equation}
Assuming that the measurement uncertainties are described by a 2-dimensional Gaussian,

\begin{center}
\hfill \parbox{7cm}{
\begin{equation}
P_i (x,y) = \frac{1}{2\pi\sigma_{x,i}\sigma_{y,i}\sqrt{1-\rho_i^2}}\exp\left[-\frac{x'^2+y'^2-2\rho_i x'y'}{2(1-\rho_i^2)}\right] \notag
\end{equation}
\begin{equation}
 \text{with} \quad x' = \frac{x-x_i}{\sigma_{x,i}}\,,~y'=\frac{y-y_i}{\sigma_{y,i}}\quad\text{and}\quad\rho=\frac{\sigma_{xy,i}^2}{\sigma_{x,i}\sigma_{y,i}}\,, \notag
\end{equation}
} \hfill
\parbox{1cm}{\begin{eqnarray} \label{eqn:2dGauss} \end{eqnarray}}
 \end{center}

where $\sigma_{x,i}^2$ and $\sigma_{y,i}^2$ are variances of the measurements $x_i$ and $y_i$, and $\sigma_{xy,i}^2=\mathrm{Cov}(x_i,y_i)$ their covariance. Finally, we specify $P(x)\propto\Theta(x_u-x)\Theta(x-x_l)$ to be constant between some upper and lower limit, $x_u$ and $x_l$. This choice mimics the implicit assumption usually made in a regression analysis, where any value of $x$ is allowed with equal probability. If the limits extend well beyond the $x$-uncertainties of the data, integral (\ref{eqn:like_abe}) can be approximated as
\begin{equation}
 \ell_i \approx c\cdot\int \mathcal{M}_{ab\eps}(y'\,|\,x')\,P_i(x_i,y_i\,|\,x',y')\, \dd x' \dd y' \label{eqn:like_abe_approx}
\end{equation}
and the constant $c$ may be omitted without loss of generality. Inserting (\ref{eqn:2dGauss}) and (\ref{eqn:mod_abe}) into (\ref{eqn:like_abe_approx}), the integral is a convolution of two Gaussians, which produces another Gaussian as the result:
\begin{center}
\hfill \parbox{7cm}{
\begin{eqnarray*}
\ell_i &=& G_{\sigma_i}(y_i-a-bx_i)~, \\ \notag
\text{where}\quad\sigma_i^2 &=& b^2\sigma_x^2-2b\sigma_{xy}^2+\sigma_y^2+\eps^2~.
\end{eqnarray*}
} \hfill
\parbox{1cm}{\begin{eqnarray} \label{eqn:like_abe_2dGauss} \end{eqnarray}}
\end{center}

If additionally one assumes no uncertainty in the $x_i$, integrating (\ref{eqn:like_abe}) over $x'$ (with $P(x')$ as defined above) is trivial, the approximation (\ref{eqn:like_abe_approx}) is not needed, and (\ref{eqn:like_abe_2dGauss}) reduces to
\begin{equation}
  \sigma_i^2 = \sigma_y^2+\eps^2~.
\end{equation}
This is the likelihood used for our adopted results, and the same as used by G09. We retain (\ref{eqn:like_abe_approx}) and (\ref{eqn:like_abe_2dGauss}) in order to explore the influence of distance and (apparent) magnitude uncertainties in \S\ref{subsec:error_treat}.

The fit is then performed by maximizing $\mathcal{L}$, or minimizing $-\ln\mathcal{L}$, under variation of the parameters $(a,b,\eps)$. Since $\mathcal{L}$ and $\ln\mathcal{L}$ are nonlinear in $(a,b,\eps)$, this requires numeric optimization, which we carry out by means of the Downhill-Simplex-Method (routine AMOEBA from the standard IDL library). The result are the best-fit parameters $(a_0,b_0,\eps_0)$ given in Tables \ref{tab:orig_full}, \ref{tab:litcomp} and \ref{tab:rels_alt}. 

\subsection{Confidence intervals}
\label{subsec:confint}
We obtain the parameters' confidence intervals by computing their respective probability distributions, $P_a(a)$, $P_b(b)$ and $P_\eps(\eps)$. These are each projections of the three-dimensional posterior probability $P(a,b,\eps\,|\,\{x\},\{y\})$ which, in turn, is related to the likelihood $\mathcal{L}$ via Bayes' theorem,
\begin{equation}
  P(a,b,\eps\,|\,\{x\},\{y\}) = c \cdot \mathcal{L}(\{x\},\{y\}\,|\,a,b,\eps) \cdot P(a,b,\eps)~,\label{eqn:post_abe}
\end{equation}
where $P(a,b,\eps)$ is the prior information and $c$ the normalization constant. The prior is chosen to be a uniform distribution in $(a,b,\ln\eps)$. Since the projected likelihood functions are well approximated by a normal ($a$ and $b$) or log-normal ($\eps$) distribution, we elect to limit the support of the prior to the intervals $[a_0-4\sigma_a,a_0+4\sigma_a]$, $[b_0-4\sigma_b,b_0+4\sigma_b]$ and $[\eps_0 e^{\sigma_\lambda^2-4\sigma_\lambda},\eps_0 e^{\sigma_\lambda^2+4\sigma_\lambda}]$, where $(\sigma_a,\sigma_b,\sigma_\lambda)$ are the standard deviations of $a$, $b$ and $\lambda\equiv\ln\eps$ as estimated from bootstrapping the best-fit values. Within these limits, we include $\approx99.99\%$ of the (projected) probability in our calculation, without covering large regions of parameter space.  We use the Kolmogorov-Smirnov test to quantify the agreement with the assumed distributions, and report the K-S-distances to be of the order of $~0.005$ ($a$ and $b$, compared to a Gaussian), and $~0.01$ ($\eps$, compared to a lognormal distribution) for all fitted data sets.

We then derive the confidence limits from the resulting cumulative distributions. We quote the central 68.3\% confidence interval, i.e. the location of the 15.9- and 84.1-precentiles relative to the respective best-fit value. For $a$ and $b$ with their normal distributions, this is the ``$1\sigma$''-interval. For the (near-)lognormal distribution followed by  $\eps$, this type of interval definition is more convenient than other common choices, such as the symmetric or the shortest interval. The parameters $(\lambda_0,\sigma_\lambda)$ of the lognormal distribution, 
\begin{equation*}
 P_\eps(\eps)=\frac{1}{\sigma_\lambda \sqrt{2\pi}}\frac{1}{\eps}\exp\left(-\frac{\ln\eps-\lambda_0}{2\sigma_\lambda}\right)
\end{equation*}
may be directly computed from the (asymmetric) interval limits, $[\eps_1,\eps_2]=[\eps_0-(\Delta\eps)_-,\eps_0+(\Delta\eps)_+]$, as given in column 7 of Table (\ref{tab:orig_full}), via
\begin{equation*}
 \lambda_0=\frac{1}{2}\ln(\eps_1\eps_2)~,~\sigma_\lambda=\frac{1}{2}\ln(\frac{\eps_2}{\eps_1})~.
\end{equation*}
Therefore, the given limits suffice to completely specify the underlying (log-normal) distribution of $\eps$, which is useful for example when deriving the \smbh\ mass function.

We note that the $\eps$-coordinate of the 3D-posterior maximum is smaller than that of $P_\eps$'s maximum (which in turn is smaller than the median and mean). The reason is that $\eps$ is positively correlated with $|a-a_0|$ and $|b-b_0|$, as it should, considering that $\eps$ ``absorbs`` the increased degree of model-data mismatch when $a$ and $b$ differ from the optimum. 

\subsection{Treatment of error in the data}
\label{subsec:error_treat}

\begin{table*}
\centering
\caption{Fitting methods and data errors}
 \begin{tabular}{*{9}{c}l}
  \toprule
   {} & luminosity & fit & $a_0$ & $\Delta a$ & $b_0$ & $\Delta b$ & ${\epsbh}_{,0}$ & $\Delta\epsbh\,(+/-)$ & error treatment \\   
   \input{table_rels_alt}
  \bottomrule
 \end{tabular}
 \tablecomments{Comparison of best-fit parameters for eqn. (\ref{eqn:linrel}) resulting from different fitting methods and error treatment. Column (1) indicates which choice of luminosity was correlated with $\mbh$ (see Table \ref{tab:lumdef}), while column (2) lists the fitting method: ``like'' (likelihood computation, which is our choice for this study), ``FITEXY'' (modified FITEXY routine) and ``like+bs'' (bootstrap resampling of the maximum-likelihood values). For ``like'' and ``FITEXY'', columns (3-8) contain the best-fit parameter values, $(a,b,\epsbh)_0$ and the central 68\%(``1-$\sigma$'')-confidence interval, $\Delta(a,b,\epsbh)$. In case of ``like+bs'', the best-fit values and intervals are replaced by the mean and standard deviation of the sample of most likely values. The last column lists which measurement errors were included in the fit. $\Delta d$ implies that distance uncertainties (see Table \ref{tab:targets}) introduce additional correlated uncertainties in both $L$ and $\mbh$, while $\Delta L_{\{\cdot\}}$ are based on rough estimates of the magnitude uncertainties: $0.1\mg$ for ellipticals' magnitudes as well as total magnitudes (all Hubble types), $0.2\mg$ for bulge magnitudes derived from decompositions, and $0.5\mg$ for the bulge magnitudes of NGC1300, NGC3245, NGC3998 and NGC4258. Printed in boldface are our adopted relations. All parameter estimates agree well within the parameter uncertainties. For details about the fitting method, please see appendix \ref{app:fitmethod}. }
\label{tab:rels_alt}
\end{table*} 

In order to fit equation (\ref{eqn:linrel}), we assume Gaussian uncertainties in $\lmbh=\log\mbh$ with standard error
\begin{equation}
 \sigma_\bullet=\frac{1}{2}\left[\log(\mbh+\Delta{\mbh}_{,+})-\log(\mbh-\Delta{\mbh}_{,-})\right]~,\label{eqn:esym}
\end{equation}
where $\mbh$ is the best-fit value, and $\Delta{\mbh}_{+/-}$ are the limits of the published $1\sigma$-confidence interval, relative to $\mbh$. That is, we symmetrize the confidence interval, while the published interval is typically asymmetric in $\log\mbh$.

To justify this symmetrization, consider that $\chi^2(\mbh)$, from which the uncertainties are typically derived, is usually not symmetric around the location of its minimum. Even if the given interval is symmetric, it may result from choosing the interval center as the best-fit value \citep[e.g.,][]{SchulzeGebhardt11}. In other words, the implied probability distribution of $\mbh$ is neither Gaussian nor symmetric to begin with, and its characterization provided in the literature ($\mbh$ and $\Delta\mbh$) is already reduced to 2 (or 3) numbers. Moreover, the limits derived from $\chi^2(\mbh)$ (e.g. $\Delta\chi^2=1$-countours) generally represent the intended confidence level (e.g. 68\%) only approximately, due to non-linearity of the model \citep[see][]{NumericalRecipies2_Fortran}. Our symmetrization therefore represents only a small approximation given the already made assumptions and the \textit{available} information on $\mbh$.

In Paper-I, we concluded that within the adopted method of image decomposition, uncertainties on the derived apparent magnitudes (bulge magnitudes in particular) cannot be assessed in a sound and reproducible manner, so that assuming no uncertainties at all might be the best option. We show now that even if some educated guess of the uncertainties is made, the effect on the fit results is comparatively small. For example, assuming normally distributed errors, and setting all $\Delta m\tot=0.1\mg$, which corresponds roughly to the differences between several methods of measuring $\mtot$ (see Paper-I, Figure 1 and Table 4), the best-fit intrinsic scatter $\eps_0$ in the $\mbh-\ltot$ relation decreases by $\approx 0.001$, that is $\approx2\%$ of the $1\sigma$-uncertainty (compare lines 6 and 10 in Table \ref{tab:rels_alt}). Likewise, $a$ and $b$ are practically unchanged. Uncertainties in bulge magnitudes, $\Delta m\sph$, should be larger than $\Delta m\tot$, although not as large as the average difference between $\lbstd$ and $\lsph$. We therefore explore the fit results from choosing $\Delta m\sph=0.5\mg$ for NGC1300, NGC3245, NGC3998 and NGC4258 (stronger residuals or larger uncertainty about the number of components) and $\Delta m\sph=0.2\mg$ for all other galaxies with a disk. These uncertainties roughly correspond to the standard deviations implied if the $\lbmax$ and $\lbmin$ are assumed as limits of a uniform distribution, and should therefore constitute a conservative (although not necessarily realistic) error estimate. Using these, the relation parameters change by $\approx20\%$ of the $1\sigma$-uncertainties (line 5 in Table \ref{tab:rels_alt}). Therefore, omitting the errors on the magnitudes does not significantly affect the accuracy our adopted results.

Finally, we \textit{can} account for uncertainty in the distance measurement: $\sigma_d=0.4\Delta(m-M)$ (see col. 4 in Table \ref{tab:targets}), which we assume to be normally distributed. On average, our error on the distance modulus is $\approx0.2\mg$, which translates to an uncertainty in 
$\lmbh$ and $\logl$ of $\approx0.08\,\mathrm{dex}$. This is about half of the typical error in $\lmbh$. Since both $\lmbh$ and $\logl$ are proportional to the distance modulus, $\sigma_d$ introduces covariance between $\lmbh$ and $\logl$. If we set the magnitude errors to zero as discussed above, the elements of the data covariance matrix are $\sigma_x^2=\sigma_d^2$, $\sigma_y^2=\sigma_\bullet^2+\sigma_d^2$ and $\sigma_{xy}^2=\sigma_d^2$ (cf. eqn. \ref{eqn:2dGauss}). Then, the likelihood of the $i$-th measurement $(x_i,y_i)$ (equation \ref{eqn:like_abe_2dGauss}) is a Gaussian with variance
\begin{equation}
 \sigma_i^2 = (1-b)^2\sigma_{d,i}^2+\sigma_{\bullet,i}^2+\eps^2~. \notag
\end{equation} 
This implies that for our relations, which have a slope $b\approx1$, the influence of $\sigma_d$ will be small. For example, assuming $\sigma_{d,i}=\sigma_{\bullet,i}/2$, $\eps=0.4\approx2\sigma_{\bullet,i}$ and a slope as low as $b=0.7$, the contribution of $\sigma_d$ to the above variance is $\approx1\%$. The contribution of $\sigma_d$ would further decrease if magnitude errors were nonzero. It vanishes altogether if $b=1$, since then the errors due to distance are ``parallel to'' the relation. In Table \ref{tab:rels_alt}, we present the relation parameters derived from including distance uncertainties (lines 4 and 9): they are very similar to our adopted relations (lines 1 and 6), as expected.

\section{Significance of parameter differences}
\label{app:pardiff}

In Tables \ref{tab:bcomp}-\ref{tab:epsortcomp}, we present the differences between our parameter estimates for selected $\mbh-L_{\{\cdot\}}$ relations as resulting from our WIRCam imaging. In these tables, the row and column headers label the kind of luminosity measurement being correlated with $\mbh$, listed in order of ascending parameter mean (which may differ slightly from the best-fit value), as given in brackets below the columns headers. Each pairwise difference is expressed as a fraction of the standard deviation of the respective difference distribution (lower-left portion of each table). In the top-right portion of each table, we supplement the corresponding confidence level (assuming a Gaussian distribution). The relations' $\epsbh$ approximately follow a log-normal distribution; hence $\log\epsbh$ and their differences are normally distributed and used to determine the standard deviation. $\epso$, the intrinsic scatter orthogonal to the correlation ridge line, is not a coordinate of the parameter grid but related to $\log\epsbh$ and the slope $b$ as $\epso=\epsbh(1+b^2)^{-1/2}$. We calculated its mean and standard deviation from bootstrap resampling. In each table, we highlight the comparison between our adopted relations ($\mbh-\lsph$ and $\mbh-\ltimp$) in boldface. The tables illustrate our main result: that the correlation of $\mbh$ with bulge luminosity is as tight as the correlation with total luminosity at the 68\%-confidence level, even when improved decompositions ($\lsph$) are applied instead of bulge/disk decompositions ($\lbstd$). 

\begin{table} 
 \begin{center}
 \caption{Relation slope}
 \input{table_bcomp}
 \tablecomments{Significances of pairwise differences between correlation slopes ($b$). For example (printed in boldface), $b\sph$ of the $\mbh-\lsph$ relation differs from $b\timp$ by 0.98 standard deviations, corresponding to the 67\%-confidence level.}
\label{tab:bcomp}
\end{center}
\end{table} 

\begin{table} 
\begin{center}
 \caption{Intrinsic relation scatter}
 \input{table_epscomp}
 \tablecomments{As Table \ref{tab:bcomp}, but for $\epsbh$. For example, ${\epsbh}\sph$ of the $\mbh-\lsph$ relation differs from ${\epsbh}\timp$ by $0.55\sigma$, i.e. at the 42\%-confidence level. All $\epsbh$ (including fits to elliptical galaxies only) agree at the $61\%$-confidence level.}
\label{tab:epscomp}
\end{center}
\end{table} 

\begin{table} 
\begin{center}
 \caption{Orthogonal intrinsic scatter}
 \input{table_epsortcomp}
 \tablecomments{As Table \ref{tab:bcomp}, but for $\epsbh$. For example, ${\epso}\sph$ of the $\mbh-\lsph$ relation differs from ${\epso}\timp$ by $0.19\sigma$, i.e. at the 15\%-confidence level. Disregarding the relation of elliptical galaxies (first row/column), all relations' $\epso$ agree at the $37\%$-confidence level.}
\label{tab:epsortcomp}
\end{center}
\end{table} 

\end{document}

%% file: Mbh-Lbt_abbrev.tex
\newcommand{\galfit}{\textsc{galfit}}
\newcommand{\gfthree}{\textsc{galfit3}}
\newcommand{\sersic}{S\'{e}rsic}
\newcommand{\iraf}{\textsc{IRAF}}

\newcommand{\fitexy}{\textsc{{FITEXY}}}
\newcommand{\twomass}{\textsc{2MASS}}      
\newcommand{\ukidss}{\textsc{UKIDSS}}      

\newcommand{\irac}{\textsc{IRAC}}
\newcommand{\smbh}{SMBH}
\newcommand{\mbh}{M_\bullet} 

\newcommand{\eps}{\epsilon} 
\newcommand{\epsbh}{\epsilon_\bullet} 
\newcommand{\epso}{\eps_\bot} 
\newcommand{\lmbh}{\tilde{\mbh}} 
\newcommand{\logl}{\tilde{L}} 
\newcommand{\dd}{\mathrm{d}} 

\newcommand{\magarcsec}{\,\mathrm{mag\,arcsec^{-2}}}
\newcommand{\mg}{\,\mathrm{mag}}

\newcommand{\msun}{\,M_\odot}
\newcommand{\lsun}{\,L_\odot}
\newcommand{\dex}{\,\mathrm{dex}}
\newcommand{\mikron}{\,\mu\mathrm{m}}

\newcommand{\std}{_\mathrm{std}}
\newcommand{\imp}{_\mathrm{imp}}
\newcommand{\bul}{_\mathrm{bul}}
\newcommand{\bstd}{_\mathrm{b,std}}

\newcommand{\bmin}{_\mathrm{b,min}}
\newcommand{\bmax}{_\mathrm{b,max}}
\newcommand{\sph}{_\mathrm{sph}}
\newcommand{\tot}{_\mathrm{tot}}
\newcommand{\ser}{_\mathrm{ser}}
\newcommand{\tstd}{_\mathrm{t,std}}
\newcommand{\timp}{_\mathrm{t,imp}}
\newcommand{\iso}{_\mathrm{24}}
\newcommand{\twom}{_\mathrm{t,2M}}
\newcommand{\mh}{_\mathrm{MH03}}

\newcommand{\ellip}{^\mathrm{(ell)}}
\newcommand{\lstd}{L\std}
\newcommand{\limp}{L\imp}
\newcommand{\lbul}{L\bul}
\newcommand{\lbstd}{L\bstd}

\newcommand{\lbmin}{L\bmin}
\newcommand{\lbmax}{L\bmax}
\newcommand{\lsph}{L\sph}
\newcommand{\ltot}{L\tot}
\newcommand{\lser}{L\ser}
\newcommand{\ltstd}{L\tstd}
\newcommand{\ltimp}{L\timp}
\newcommand{\liso}{L\iso}
\newcommand{\ltwom}{L\twom}
\newcommand{\lmh}{L\mh}

\newcommand{\mtot}{m\tot}

\newcommand{\Mbul}{M\bul}

%% file: table_targets.tex
CygA & E & 238.0 & 26.5 & 6.4~/~6.4 & 13.2 & 19.6 & 2-1 \\
IC1459 & E3 & 28.4 & 24.4 & 2.8~/~2.8 & 3.71 & 3.98 & 1-2 \\
IC4296 & E & 50.8 & 13.4 & 2.1~/~1.9 & 7.39 & 9.33 & 7-3 \\
NGC0221 & cE2 & 0.791 & 0.028 & 0.006~/~0.006 & 0.0133 & 0.124 & 1-4 \\
NGC0821 & E6 & 23.4 & 1.65 & 0.74~/~0.74 & 1.01 & 2.85 & 1-5 \\
NGC1023 & S0 & 11.1 & 0.42 & 0.04~/~0.04 & 0.944 & 1.92 & 1-6 \\
NGC1300 & SBbc & 19.0 & 0.67 & 0.64~/~0.32 & 0.889 & 6.42 & 2-7 \\
NGC1399 & E1pec & 21.2 & 5.1 & 0.7~/~0.7 & 3.89 & 5.72 & 8-8 \\
NGC2748 & SAbc & 24.0 & 0.45 & 0.36~/~0.37 & 0.402 & 1.87 & 2-7 \\
NGC2778 & S0 & 22.3 & 0.129 & 0.1~/~0.1 & 0.174 & 1.07 & 1-5 \\
NGC2787 & SB(r)0 & 7.28 & 1.04 & 0.36~/~0.64 & 0.149 & 0.683 & 1-9 \\
NGC3115 & S0 & 9.42 & 8.85 & 2.8~/~2.8 & 1.03 & 1.63 & 1-10 \\
NGC3227 & SAB(s)pec & 16.8 & 0.13 & 0.06~/~0.06 & 0.586 & 2.55 & 2-11 \\
NGC3245 & SA0(r)? & 20.3 & 2.04 & 0.49~/~0.49 & 0.691 & 1.54 & 1-12 \\
NGC3377 & E5 & 10.9 & 1.77 & 0.93~/~0.93 & 0.312 & 1.47 & 1-5 \\
NGC3379 & E1 & 10.3 & 4.0 & 1.0~/~1.0 & 0.980 & 2.39 & 1-13 \\
NGC3384 & SB(s)0- & 11.3 & 0.106 & 0.048~/~0.048 & 0.539 & 1.32 & 1-5 \\
NGC3608 & E2 & 22.3 & 4.55 & 0.97~/~0.97 & 0.854 & 2.67 & 1-5 \\
NGC3998 & SA(r)0 & 13.9 & 8.2 & 2.0~/~1.9 & 0.481 & 0.918 & 1-14 \\
NGC4258 & SAB(s)bc & 7.21 & 0.39 & 0.01~/~0.01 & 0.768 & 3.03 & 4-15 \\
NGC4261 & E2 & 30.8 & 5.02 & 1.0~/~1.0 & 3.69 & 6.51 & 1-16 \\
NGC4291 & E3 & 25.5 & 9.37 & 3.0~/~3.0 & 0.728 & 1.78 & 1-5 \\
NGC4342 & S0 & 13.3 & 2.7 & 1.5~/~1.0 & 0.101 & 0.282 & 2-17 \\
NGC4374 & E1 & 18.5 & 9.27 & 0.98~/~0.87 & 3.34 & 5.25 & 5-18 \\
NGC4473 & E5 & 15.3 & 0.899 & 0.45~/~0.45 & 0.734 & 1.58 & 5-5 \\
NGC4486 & E0pec & 16.7 & 61.5 & 3.7~/~3.7 & 4.54 & 7.18 & 5-19 \\
NGC4564 & S0 & 15.9 & 0.88 & 0.24~/~0.24 & 0.369 & 1.05 & 5-5 \\
NGC4649 & E2 & 16.4 & 47.1 & 10.0~/~10.0 & 3.94 & 5.16 & 5-20 \\
NGC4697 & E6 & 12.5 & 2.02 & 0.51~/~0.51 & 1.41 & 3.82 & 1-5 \\
NGC5252 & S0 & 97.3 & 10.0 & 15.0~/~4.7 & 3.20 & 3.97 & 2-21 \\
NGC5845 & E* & 25.2 & 4.75 & 1.5~/~1.5 & 0.331 & 0.479 & 1-5 \\
NGC6251 & E & 107.0 & 6.0 & 2.0~/~2.0 & 7.88 & 9.88 & 2-22 \\
NGC7052 & E & 67.6 & 3.7 & 2.6~/~1.5 & 4.80 & 7.60 & 2-23 \\
NGC7457 & SA(rs)0- & 12.9 & 0.092 & 0.055~/~0.055 & 0.205 & 1.86 & 1-5 \\
PGC49940 & E & 153.0 & 37.6 & 4.3~/~5.2 & 7.67 & 12.3 & 2-3 \\

%% file: table_loglum.tex
CygA & 1 & $9.42\pm0.11$ & 12.01  & 12.12  & 12.10  & 12.10  & \bf 12.10  & 12.10  & 12.10  & 12.10  & \bf 12.10  & 12.22  \\
IC1459 & 1 & $9.39\pm0.05$ & 11.50  & 11.57  & 11.65  & 11.65  & \bf 11.71  & 11.65  & 11.71  & 11.71  & \bf 11.71  & 11.65  \\
IC4296 & 1 & $9.13\pm0.06$ & 11.73  & 11.87  & 11.87  & 11.87  & \bf 12.05  & 11.87  & 12.05  & 12.05  & \bf 12.05  & -- \\
NGC0221 & 1 & $6.45\pm0.09$ & 9.08   & 9.12   & 9.10   & 9.10   & \bf 9.10   & 9.10   & 9.10   & 9.10   & \bf 9.10   & 9.22   \\
NGC0821 & 2 & $8.22\pm0.21$ & 10.91  & 11.01  & 11.10  & 11.11  & \bf 11.01  & 11.11  & 10.29  & 10.99  & \bf 10.99  & 11.21  \\
NGC1023 & 2 & $7.62\pm0.04$ & 10.92  & 10.98  & 11.06  & 10.97  & \bf 10.97  & 10.65  & 10.48  & 10.59  & \bf 10.48  & 10.69  \\
NGC1300 & 3 & $7.83\pm0.29$ & 10.85  & 10.95  & 10.74  & 10.95  & \bf 10.97  & 9.92   & 10.05  & 10.38  & \bf 10.05  & -- \\
NGC1399 & 1 & $8.71\pm0.06$ & 11.44  & 11.59  & 11.59  & 11.59  & \bf 11.80  & 11.59  & 11.80  & 11.80  & \bf 11.80  & -- \\
NGC2748 & 3 & $7.65\pm0.50$ & 10.59  & 10.60  & 10.67  & 10.61  & \bf 10.61  & 10.04  & 10.15  & 10.15  & \bf 10.15  & -- \\
NGC2778 & 2 & $7.11\pm0.46$ & 10.21  & 10.24  & 10.43  & 10.24  & \bf 10.24  & 9.94   & 9.83   & 9.85   & \bf 9.83   & 10.49  \\
NGC2787 & 2 & $8.02\pm0.27$ & 10.15  & 10.17  & 10.22  & 10.16  & \bf 10.17  & 9.73   & 9.98   & 10.05  & \bf 9.98   & 9.81   \\
NGC3115 & 2 & $8.95\pm0.14$ & 10.91  & 11.01  & 11.04  & 11.08  & \bf 11.01  & 11.05  & 10.09  & 10.93  & \bf 10.93  & 11.05  \\
NGC3227 & 3 & $7.11\pm0.22$ & 10.71  & 10.77  & 11.22  & 10.76  & \bf 10.76  & 10.15  & 9.95   & 10.12  & \bf 9.95   & -- \\
NGC3245 & 2 & $8.31\pm0.11$ & 10.79  & 10.84  & 10.99  & 10.82  & \bf 10.83  & 10.35  & 10.13  & 10.35  & \bf 10.13  & 10.61  \\
NGC3377 & 2 & $8.25\pm0.25$ & 10.42  & 10.49  & 10.65  & 10.55  & \bf 10.50  & 10.54  & 10.05  & 10.39  & \bf 10.38  & 10.73  \\
NGC3379 & 1 & $8.60\pm0.11$ & 10.83  & 10.99  & 11.03  & 11.03  & \bf 11.13  & 11.03  & 11.13  & 11.13  & \bf 11.13  & 10.97  \\
NGC3384 & 2 & $7.03\pm0.21$ & 10.72  & 10.73  & 10.72  & 10.75  & \bf 10.73  & 10.38  & 10.18  & 10.45  & \bf 10.18  & 10.33  \\
NGC3608 & 1 & $8.66\pm0.09$ & 10.77  & 10.93  & 11.05  & 11.05  & \bf 11.05  & 11.05  & 11.05  & 11.05  & \bf 11.05  & 10.93  \\
NGC3998 & 2 & $8.91\pm0.11$ & 10.65  & 10.68  & 10.72  & 10.68  & \bf 10.67  & 10.40  & 9.94   & 10.34  & \bf 9.94   & -- \\
NGC4258 & 3 & $7.59\pm0.01$ & 10.85  & 10.89  & 11.09  & 10.93  & \bf 10.90  & 10.52  & 9.33   & 9.96   & \bf 9.33   & 10.27  \\
NGC4261 & 1 & $8.70\pm0.09$ & 11.39  & 11.57  & 11.51  & 11.51  & \bf 11.63  & 11.51  & 11.63  & 11.63  & \bf 11.63  & 11.53  \\
NGC4291 & 1 & $8.97\pm0.14$ & 10.76  & 10.86  & 10.89  & 10.89  & \bf 10.94  & 10.89  & 10.94  & 10.94  & \bf 10.94  & 10.85  \\
NGC4342 & 2 & $8.43\pm0.20$ & 9.96   & 10.00  & 9.93   & 9.97   & \bf 9.98   & 9.68   & 9.47   & 9.65   & \bf 9.65   & 9.73   \\
NGC4374 & 1 & $8.97\pm0.04$ & 11.37  & 11.52  & 11.59  & 11.59  & \bf 11.69  & 11.59  & 11.69  & 11.69  & \bf 11.69  & 11.60  \\
NGC4473 & 1 & $7.95\pm0.24$ & 10.82  & 10.87  & 10.89  & 10.89  & \bf 10.89  & 10.89  & 10.89  & 10.89  & \bf 10.89  & 10.81  \\
NGC4486 & 1 & $9.79\pm0.03$ & 11.44  & 11.66  & 11.57  & 11.57  & \bf 11.75  & 11.57  & 11.75  & 11.75  & \bf 11.75  & 11.58  \\
NGC4564 & 2 & $7.94\pm0.12$ & 10.55  & 10.57  & 10.60  & 10.58  & \bf 10.58  & 10.43  & 10.43  & 10.43  & \bf 10.43  & 10.72  \\
NGC4649 & 1 & $9.67\pm0.10$ & 11.45  & 11.60  & 11.53  & 11.53  & \bf 11.67  & 11.53  & 11.67  & 11.67  & \bf 11.67  & 11.61  \\
NGC4697 & 2 & $8.31\pm0.11$ & 10.96  & 11.15  & 11.27  & 11.28  & \bf 11.19  & 11.27  & 10.08  & 11.15  & \bf 11.14  & 11.21  \\
NGC5252 & 2 & $9.00\pm0.34$ & 11.38  & 11.50  & 11.63  & 11.38  & \bf 11.54  & 10.89  & 11.39  & 11.48  & \bf 11.39  & 11.56  \\
NGC5845 & 1 & $8.68\pm0.14$ & 10.48  & 10.52  & 10.48  & 10.48  & \bf 10.48  & 10.48  & 10.48  & 10.48  & \bf 10.48  & 10.49  \\
NGC6251 & 1 & $8.78\pm0.15$ & 11.77  & 11.90  & 11.90  & 11.90  & \bf 11.90  & 11.90  & 11.90  & 11.90  & \bf 11.90  & 11.95  \\
NGC7052 & 1 & $8.57\pm0.23$ & 11.56  & 11.68  & 11.67  & 11.67  & \bf 11.67  & 11.67  & 11.67  & 11.67  & \bf 11.67  & 11.70  \\
NGC7457 & 2 & $6.96\pm0.30$ & 10.26  & 10.31  & 10.63  & 10.35  & \bf 10.29  & 10.05  & 9.15   & 9.60   & \bf 9.15   & 10.01  \\
PGC49940 & 1 & $9.58\pm0.06$ & 11.70  & 11.88  & 11.88  & 11.88  & \bf 11.88  & 11.88  & 11.88  & 11.88  & \bf 11.88  & -- \\

%% file: table_orig_full.tex
{} & (1) & (2) & (3) & (4) & (5) & (6) & (7) & (8) & (9) & (10) \\
\hline \\
$(1)$ & $\lbstd$ & $8.53$ & $\pm0.09$ & $0.86$ & $\pm0.12$ & $0.48$ & $+0.10 / -0.04$ & $0.36$ & $0.04$ & $0.78$ \\
$(2)$ & $\lbmin$ & $8.61$ & $\pm0.10$ & $0.70$ & $\pm0.10$ & $0.49$ & $+0.10 / -0.04$ & $0.40$ & $0.05$ & $0.78$ \\
$(3)$ & $\lbmax$ & $8.51$ & $\pm0.09$ & $0.83$ & $\pm0.11$ & $0.45$ & $+0.09 / -0.04$ & $0.34$ & $0.05$ & $0.82$ \\
$\boldsymbol{(4)} $ & $\boldsymbol{\lsph} $ & $\boldsymbol{8.56} $ & $\boldsymbol{\pm0.09} $ & $\boldsymbol{0.75} $ & $\boldsymbol{\pm0.10} $ & $\boldsymbol{0.46} $ & $\boldsymbol{+0.09 / -0.04} $ & $\boldsymbol{0.36} $ & $\boldsymbol{0.05} $ & $\boldsymbol{0.81} $ \\
\hline
$(5)$ & $\lser$ & $8.35$ & $\pm0.11$ & $0.90$ & $\pm0.17$ & $0.57$ & $+0.11 / -0.05$ & $0.42$ & $0.06$ & $0.68$ \\
\hline
$(6)$ & $\ltstd$ & $8.39$ & $\pm0.10$ & $0.95$ & $\pm0.16$ & $0.52$ & $+0.10 / -0.05$ & $0.38$ & $0.05$ & $0.74$ \\
$\boldsymbol{(7)} $ & $\boldsymbol{\ltimp} $ & $\boldsymbol{8.36} $ & $\boldsymbol{\pm0.10} $ & $\boldsymbol{0.92} $ & $\boldsymbol{\pm0.14} $ & $\boldsymbol{0.51} $ & $\boldsymbol{+0.10 / -0.04} $ & $\boldsymbol{0.37} $ & $\boldsymbol{0.05} $ & $\boldsymbol{0.75} $ \\
$(8)$ & $\liso$ & $8.40$ & $\pm0.10$ & $0.96$ & $\pm0.15$ & $0.51$ & $+0.10 / -0.05$ & $0.37$ & $0.05$ & $0.75$ \\
\hline
$(9)$ & $\lstd^{(ell)}$ & $8.56$ & $\pm0.12$ & $0.90$ & $\pm0.16$ & $0.40$ & $+0.14 / -0.04$ & $0.30$ & $0.05$ & $0.82$ \\
$(10)$ & $\limp^{(ell)}$ & $8.50$ & $\pm0.13$ & $0.88$ & $\pm0.15$ & $0.39$ & $+0.13 / -0.04$ & $0.30$ & $0.04$ & $0.83$ \\
$(11)$ & $\liso^{(ell)}$ & $8.56$ & $\pm0.12$ & $0.91$ & $\pm0.16$ & $0.39$ & $+0.13 / -0.04$ & $0.29$ & $0.04$ & $0.83$ \\

%% file: table_litcomp.tex
 & (1) & (2) & (3) & (4) & (5) & (6) & (7) & (8) & (9) & (10) & (11) & (12) \\
\midrule
(1) & MH03 & \twomass\ $K$ & all & 37 & AB96 & 8.20 & 0.10 & 1.21 & 0.13 & 0.51 & -- &  \\
(2) & -- & \twomass\ $K$ & all & 37 & like & 8.16 & 0.10 & 0.91 & 0.15 & 0.53 & 0.10/0.05 & $\mbh,L$ as in MH03 \\
(3) & MH03 & \twomass\ $K$ & rel & 27 & AB96 & 8.32 & 0.07 & 1.13 & 0.12 & 0.31 & -- & only secure $\mbh$ \\
(4) & -- & \twomass\ $K$ & rel & 27 & like & 8.30 & 0.08 & 1.03 & 0.11 & 0.33 & 0.08/0.03 & $\mbh,L$ as in MH03 \\
(5) & G07 & \twomass\ $K$ & rel & 22 & T02 & 8.17 & 0.08 & 0.93 & 0.10 & 0.30 & 0.03/0.05\footnote{probably upper and lower error bar are confused in Table 4 of G07 (V12 and this work find the upper error on $\eps$ to be larger than the lower error bar)} & 5 galaxies removed\footnote{with respect to MH03's sample of reliable $\mbh$; for details about the removals as well as applied updates of some $\mbh$, $\lbul$ and distances, see G07} \\
(6) & S11 & \irac\ $3.6\mu\mathrm{m}$ & res & 48 & like\footnote{LINMIX\_ERR (Kelly 2007) as stated in S11} & 8.19 & 0.06 & 0.93 & 0.10 & 0.38 & 0.05 & no pseudobulges \\
(7) & V12 & \ukidss\ $K$ & all & 25 & T02 & 8.38 & 0.20\footnote{the larger error in $a$ is a result of correlation with $b$, in turn due to V12's choice of magnitude offset ($-18\mg$ instead of the magnitude mean)} & 0.88 & 0.06 & 0.52 & 0.10/0.06 &  \\
(8) & V12 & \ukidss\ $K$ & res & 19 & T02 & 8.71 & 0.25 & 0.90 & 0.08 & 0.40 & 0.09/0.06 & no N4342,M87,barred\footnote{NGC4342 has been removed because it was deemed an outlier of the $\mbh-\lbul$ relation, M87 because of poor image (sky background) quality, and barred galaxies as identified in V12} \\ \hline
(9) & -- & \twomass\ $K$ & $\cap$ & 28 & like & 8.29 & 0.10 & 0.91 & 0.14 & 0.45 & 0.11/0.04 & $\mbh,L$ from MH03$^\parallel$ \\
(10) & -- & \twomass\ $K$ & $\cap$ & 28 & like & 8.47 & 0.10 & 0.89 & 0.14 & 0.47 & 0.11/0.04 & $L$ from MH03$^\parallel$ \\
(11) & -- & WIRCam $K$ & $\cap$ & 28 & like & 8.39 & 0.09 & 0.80 & 0.10 & 0.41 & 0.10/0.04 & $\mbh$ from MH03\footnote{values rescaled to our adopted distances} \\
(12) & -- & WIRCam $K$ & $\cap$ & 28 & like & 8.56 & 0.09 & 0.79 & 0.11 & 0.42 & 0.10/0.04 &  \\ \hline
\bf (13) & \bf -- & \bf WIRCam $\boldsymbol{K}$ & \bf all & \bf 35 & \bf like & \bf 8.56 & \bf 0.09 & \bf 0.75 & \bf 0.10 & \bf 0.46 & \bf 0.09/0.04 & \bf  \\

%% file: table_rels_alt.tex
{} & (1) & (2) & (3) & (4) & (5) & (6) & (7) & (8) & (9) \\
\hline \\
\bf (1) & \bf $\boldsymbol{\lsph}$ & \bf like & \bf 8.56 & \bf 0.09 & \bf 0.75 & \bf 0.10 & \bf 0.46 & \bf 0.09 / 0.04 & \bf $\boldsymbol{\Delta\mbh}$ \\
(2) & $\lsph$ & like+bs & 8.56 & 0.08 & 0.75 & 0.10 & 0.43 & 0.06 & $\Delta\mbh$ \\
(3) & $\lsph$ & FITEXY & 8.56 & 0.09 & 0.75 & 0.10 & 0.46 & -- & $\Delta\mbh$ \\
(4) & $\lsph$ & like & 8.56 & 0.09 & 0.75 & 0.10 & 0.46 & 0.09 / 0.04 & $\Delta\mbh$ and $\Delta d$ \\
(5) & $\lsph$ & like & 8.55 & 0.09 & 0.76 & 0.10 & 0.45 & 0.09 / 0.04 & $\Delta\mbh$ and $\Delta\lsph$ \\ \hline
\bf (6) & \bf $\boldsymbol{\ltimp}$ & \bf like & \bf 8.36 & \bf 0.10 & \bf 0.92 & \bf 0.14 & \bf 0.51 & \bf 0.10 / 0.04 & \bf $\boldsymbol{\Delta\mbh}$ \\
(7) & $\ltimp$ & like+bs & 8.36 & 0.10 & 0.93 & 0.13 & 0.49 & 0.06 & $\Delta\mbh$ \\
(8) & $\ltimp$ & FITEXY & 8.36 & 0.09 & 0.92 & 0.14 & 0.52 & -- & $\Delta\mbh$ \\
(9) & $\ltimp$ & like & 8.36 & 0.10 & 0.92 & 0.14 & 0.51 & 0.10 / 0.04 & $\Delta\mbh$ and $\Delta d$ \\
(10) & $\ltimp$ & like & 8.36 & 0.10 & 0.92 & 0.14 & 0.51 & 0.10 / 0.04 & $\Delta\mbh$ and $\Delta\ltimp$ \\

%% file: table_bcomp.tex
 \begin{tabular}{l|*{5}c}
 & $\lbmin$ & $\lsph$ & $\lbstd$ & $\limp^{(ell)}$ & $\ltimp$ \\
\multicolumn{1}{r|}{$\langle b \rangle$} & (0.699) & (0.751) & (0.857) & (0.876) & (0.923) \\
\hline
$\lbmin$ & -- & 28.4\% & 67.6\% & 66.2\% & 79.5\% \\
$\lsph$ & 0.364 & -- & 49.7\% & 50.5\% & \bf 67.4\% \\
$\lbstd$ & 0.986 & 0.669 & -- &  7.8\% & 27.3\% \\
$\limp^{(ell)}$ & 0.958 & 0.683 & 0.097 & -- & 17.7\% \\
$\ltimp$ & 1.268 & \bf 0.983 & 0.350 & 0.223 & -- \\
\end{tabular}

%% file: table_epscomp.tex
 \begin{tabular}{l|*{5}c}
 & $\limp^{(ell)}$ & $\lsph$ & $\lbstd$ & $\lbmin$ & $\ltimp$ \\
\multicolumn{1}{r|}{$\langle \epsbh \rangle$} & (0.432) & (0.477) & (0.502) & (0.509) & (0.530) \\
\hline
$\limp^{(ell)}$ & -- & 32.3\% & 47.3\% & 51.2\% & 61.3\% \\
$\lsph$ & 0.417 & -- & 20.8\% & 26.5\% & \bf 41.5\% \\
$\lbstd$ & 0.632 & 0.263 & -- &  6.0\% & 22.3\% \\
$\lbmin$ & 0.694 & 0.339 & 0.075 & -- & 16.5\% \\
$\ltimp$ & 0.864 & \bf 0.546 & 0.283 & 0.208 & -- \\
\end{tabular}

%% file: table_epsortcomp.tex
 \begin{tabular}{l|*{5}c}
 & $\limp^{(ell)}$ & $\lsph$ & $\lbstd$ & $\ltimp$ & $\lbmin$ \\
\multicolumn{1}{r|}{$\langle \epso \rangle$} & (0.280) & (0.346) & (0.350) & (0.358) & (0.380) \\
\hline
$\limp^{(ell)}$ & -- & 68.9\% & 74.1\% & 78.7\% & 87.1\% \\
$\lsph$ & 1.014 & -- &  4.3\% & \bf 14.7\% & 37.0\% \\
$\lbstd$ & 1.130 & 0.054 & -- & 11.1\% & 34.8\% \\
$\ltimp$ & 1.246 & \bf 0.186 & 0.139 & -- & 24.4\% \\
$\lbmin$ & 1.518 & 0.481 & 0.451 & 0.311 & -- \\
\end{tabular}